\documentclass[a4paper,fleqn,usenatbib,useAMS]{mnras}
\usepackage{newtxtext,newtxmath}
\usepackage[T1]{fontenc}
\usepackage{ae,aecompl}
\usepackage{graphicx,color}
\usepackage{amsmath,amsfonts,amssymb,cancel,ulem} 
\usepackage{times}
\usepackage{hyperref}
\usepackage{caption}
\hypersetup{
    breaklinks=true,
    colorlinks=true,
    linkcolor={blue},
    citecolor={magenta},
    urlcolor={blue}
}

\def\apj{Astrophys.\ J.}
\def\apss{Astrophys.\ Space Sci.}
\def\an{Astron.\ Nachr.}
\def\mnras{Mon.\ Not.\ R.\ Astron.\ Soc.}
\def\aap{Astron.\ Astrophys.}
\def\apjl{Astrophys.\ J.\ Lett.}

\def\jfm{J.\ Fluid Mech.}
\def\jcap{J.\ Cosmol.\ Astropart.\ Phys.}
\def\physrep{Phys.\ Rep.}

\def\prl{Phys.\ Rev.\ Lett.}
\def\prd{Phys.\ Rev.\ D}
\def\apjs{ApJS}

\def\nat{Nature}

\def\ssr{SSRv}

\def\araa{Ann.\ Rev.\ Astron.\ Astrophys.}

\def\jpp{J.\ Plasma.\ Phys.}

\newcommand\sal{Sov.\ Astron.\ Lett.}

\newcommand\zhetp{Zh.\ Eksper.\ Teor.\ Fiz.}

\newcommand{\Lar}{r_{\text{L}}}    

\newcommand{\cra}{_\mathrm{cr}}
\newcommand{\Al}{_\mathrm{A}}
\newcommand{\Rm}{R_\mathrm{m}} 
\newcommand{\Rmc}{R_\mathrm{m, c}}    

\renewcommand{\vec}[1]{\mathbf{#1}}	
\newcommand{\dd}{\mathrm{d}}        
\newcommand{\e}{\mathrm{e}}        

\newcommand\deriv[2]{\displaystyle\frac{\partial #1}{\partial #2} }
\newcommand\oderiv[2]{\displaystyle\frac{\dd #1}{\dd #2} }
\newcommand{\dsp}{\displaystyle}
\newcommand\Eq[1]{Eq.~(\ref{#1})} 
\newcommand\Fig[1]{Fig.~\ref{#1}}

\newcommand{\cm}{\,{\rm cm}}    
\newcommand{\pc}{\,{\rm pc}}     
\newcommand{\kpc}{\,{\rm kpc}}  

\newcommand{\s}{\,{\rm s}}      
\newcommand{\yr}{\,{\rm yr}}    

\newcommand{\mkG}{\,\mu{\rm G}} 

\newcommand{\GeV}{\,{\rm GeV}}  

\newcommand{\tdiff}{\,t_{\rm d}}
\newcommand{\Brms}{\,B_{\rm rms}}
\newcommand{\brms}{\,b_{\rm rms}}
\newcommand{\Beq}{\,B_{\rm eq}}

\newcommand{\ncr}{n_{\rm cr}}
\newcommand{\ncravg}{\langle n_{\rm cr}\rangle}

\newcommand{\Alfven}{Alfv\'en } 

\title[Cosmic rays and magnetic fields]{Relative distribution of cosmic rays and magnetic fields}

\author[Amit Seta et al.]{Amit Seta,$^1$\thanks{\href{mailto:a.seta1@ncl.ac.uk}{a.seta1@ncl.ac.uk}}
Anvar Shukurov,$^1$ Toby S. Wood,$^1$   Paul J. Bushby$^1$ and
Andrew P. Snodin$^2$
\\
$^1$School of Mathematics and Statistics, Newcastle University, Newcastle Upon Tyne, NE1 7RU, UK \\
$^2$Department of Mathematics, Faculty of Applied Science, King Mongkut's University of Technology North Bangkok,
Bangkok 10800, Thailand
}

\begin{document}
\label{firstpage}
\pagerange{\pageref{firstpage}--\pageref{lastpage}}
\maketitle
\begin{abstract}
Synchrotron radiation from cosmic rays is a key observational probe of the galactic magnetic field.
Interpreting synchrotron emission data requires knowledge of the cosmic ray number density,
which is often assumed to be in energy equipartition (or otherwise tightly correlated)
with the magnetic field energy.
However, there is no compelling observational or theoretical reason to expect such tight correlation to hold across all scales.
We use test particle simulations, tracing the propagation of charged particles (protons) through a random magnetic field,
to study the cosmic ray distribution at scales comparable to
the correlation scale of the turbulent flow in the interstellar medium ($\simeq 100\pc$ in spiral galaxies).
In these simulations, we find that there is no spatial correlation between the cosmic ray number density and the magnetic field energy density.
In fact, their distributions are approximately
statistically independent.
We find that low-energy cosmic rays can become trapped between magnetic mirrors,
whose location depends more on the structure of the field lines than on the field strength.
\end{abstract}

\begin{keywords}
cosmic rays -- ISM:magnetic fields -- scattering -- dynamo -- MHD -- radio continuum:ISM
\end{keywords}

\section{Introduction} \label{intro}
Synchrotron emission is the main source of information about the nonthermal components of the interstellar
and intergalactic medium (ISM and IGM). Its interpretation depends
crucially on the relative distribution of magnetic fields and
cosmic ray electrons. Cosmic rays tend to gyrate around magnetic field lines, so it is natural to expect some
correlation between the cosmic ray and magnetic energy densities. 
On sufficiently long length- and time-scales, the cosmic ray distribution may achieve energy equipartition \citep{B56},
or pressure balance, with the magnetic field \citep[see review by][]{Beck2005}.
Although the assumption of a tight, point-wise correlation between cosmic 
rays and magnetic fields across a wide range of scales lacks a compelling justification, it is often used in
interpretations of synchrotron observations, regardless of the spatial resolution.
Most of the energy of cosmic rays is carried by protons and heavier particles; therefore,
such interpretations rely on an additional assumption that
relativistic electrons are distributed similarly to the heavier cosmic ray particles.

The diffusivity of $5 \GeV$ cosmic rays in a $5 \mkG$ magnetic field is
believed to be about $10^{28}$--$10^{29}\cm^2\s^{-1}$, implying
a diffusion length
of the order of $1\kpc$ over the time scale of cosmic ray confinement in galaxies, $10^6\yr$ \citep{BBDP90}.
This suggests that the cosmic ray distribution will be significantly more uniform than that
of the interstellar magnetic field, which varies strongly on scales smaller than $0.1\kpc$ \citep{RSS89,Beck1996}.
However, cosmic rays are more tightly confined to field lines where the field is strongest, and so we might still expect some positive correlation 
between cosmic ray density and magnetic energy on small scales. On the other hand, overall pressure balance can, equally
plausibly, lead to local anticorrelation between cosmic ray and magnetic energy densities \citep{Beck2003}.
Indeed, from a comparison of the observed and modelled magnitude of the fluctuations in synchrotron intensity in the Milky Way and nearby spiral galaxies, \citet{Stepanov2014} suggested 
that cosmic ray electrons and interstellar magnetic fields are slightly anticorrelated at scales $\lesssim 100\pc$.

Large-scale simulations of cosmic ray propagation in the advection-diffusion
approximation rely on various (often crude) parameterisations of the diffusion tensor and its dependence
on the magnetic field. 
Using a fluid description of cosmic rays in nonlinear dynamo simulations, assuming anisotropic (but constant) diffusion coefficients,
\cite{SBMS06} found no correlation between magnetic field and cosmic ray energy densities.
Such simulations may be appropriate at scales exceeding the diffusion length (of order
$1\kpc$), but not at the smaller scales that are relevant in the present study. 
For a typical cosmic ray proton, with energy $1$--$10 \GeV$,
in a $5 \mkG$ magnetic field, the Larmor radius
is about $10^{-6}$--$10^{-5} \pc$, which is much
smaller than the correlation length of magnetic field ($50$--$100 \pc$) or 
the diffusion length of cosmic rays (which is of the order of $1 \kpc$). Thus, the small-scale structure
of the magnetic field will be particularly important for the propagation of cosmic rays in this energy range.

In this paper, we use test particle simulations to explore, in detail, the
spatial distribution of cosmic ray particles propagating in a random magnetic field.
Whereas previous test particle simulations
have been mostly concerned with the calculation of the cosmic ray diffusion coefficient,
we primarily consider the spatial distribution of cosmic rays and its relation to the magnetic field.
Our model is kinematic,
i.e., we only consider the effect of the magnetic field on the cosmic rays
and neglect any effects of cosmic ray pressure on the gas flow and hence on the magnetic field.
In Section~\ref{RMF} we describe the numerical model for random magnetic fields and 
in Section~\ref{crpropg} we present our model for simulating cosmic ray propagation.
Our results on cosmic ray density and its relation to the magnetic field are presented in Section~\ref{results}. We conclude in Section~\ref{conclusions} and suggest further avenues for study.

\section{Implementation of random magnetic fields}\label{RMF}
The propagation of cosmic rays is sensitive to rather subtle details of the magnetic field in which they move.
The spectrum provides a complete statistical description of a Gaussian random 
magnetic field, so it must also determine the corresponding cosmic ray diffusivity for a given energy of particle. 
The propagation of cosmic rays in an isotropic Gaussian random magnetic field (for which the probability distribution function of each vector component is Gaussian) 
has been the subject of many studies \citep{BBDP90,MichalekOstrowski97,GJ1999,Casse_et_al2002,Sch02,Parizot2004,
CandiaRoulet2004,DBS2007,GAP08,Sh09,Plotnikov_et_al2011,HMR14,Snodin_et_al2016,Subedi+17}.
However, radio \citep{Gaenslar2011,HS13},
submillimeter \citep{Z+15}
and neutral hydrogen \citep{HT05,KK2016} 
observations
suggest
that the magnetic field in the ISM is
strongly non-Gaussian, spatially intermittent, and 
filamentary.
Such an intermittent field is
also expected theoretically,
as a result of turbulent dynamo action 
\citep{Wilkin2007} and 
random shock compression \citep{BT85,ByT87,B88}.
The magnetic field generated by dynamo action in galaxy clusters is also likely to be intermittent
\citep{RSS89,SSH06}.
The presence of magnetic intermittency can significantly affect the propagation of cosmic rays
\citep{SSSBW17}. In non-Gaussian fields, the separation and size of magnetic structures may play an important role 
\citep{SSSBW17}, especially for low energy particles. In this section, we describe the magnetic fields that
are used in our analysis of cosmic ray propagation. 
A discussion of the structure of small-scale interstellar magnetic field can be found in Appendix~\ref{finescale}. 
Throughout the text, the small-scale or fluctuating field is represented by $\vec{b}$, the large-scale (or mean) field
by $\vec{B_0}$ and the total field by $\vec{B}=\vec{b}+\vec{B_0}$.

\subsection{Magnetic fields generated by a random flow}\label{MFGRF}

For our numerical study, we use a random magnetic field produced by kinematic fluctuation (small-scale) dynamo action.
Using a triply-periodic cubic box of length $L$, with $512^3$ grid points, we solve the induction equation
\begin{equation}\label{indeqn}
\deriv{\vec{b}}{t} = \nabla \times (\vec{u} \times \vec{b}) + \eta \nabla^2 \vec{b}, 
\end{equation} 
where $\vec{u}(\vec{x},t)$ is a prescribed velocity field and $\eta$ is the magnetic diffusivity,
which we take to be constant.
To ensure $\nabla \cdot \vec{b} = 0$, we numerically solve for the corresponding magnetic vector potential.
We define the magnetic Reynolds number to be $\Rm = l_0 u_0/\eta$,
where $l_{0}$ is the outer scale of the turbulent velocity flow and $u_{0}$ is the
root-mean-square (rms) velocity.
Dynamo action occurs, i.e., the magnetic field grows exponentially,
if $\Rm$ exceeds a critical magnetic Reynolds number, $\Rmc$, whose value depends on the velocity field.

\begin{figure*} \centering 
      \includegraphics[width=\columnwidth]{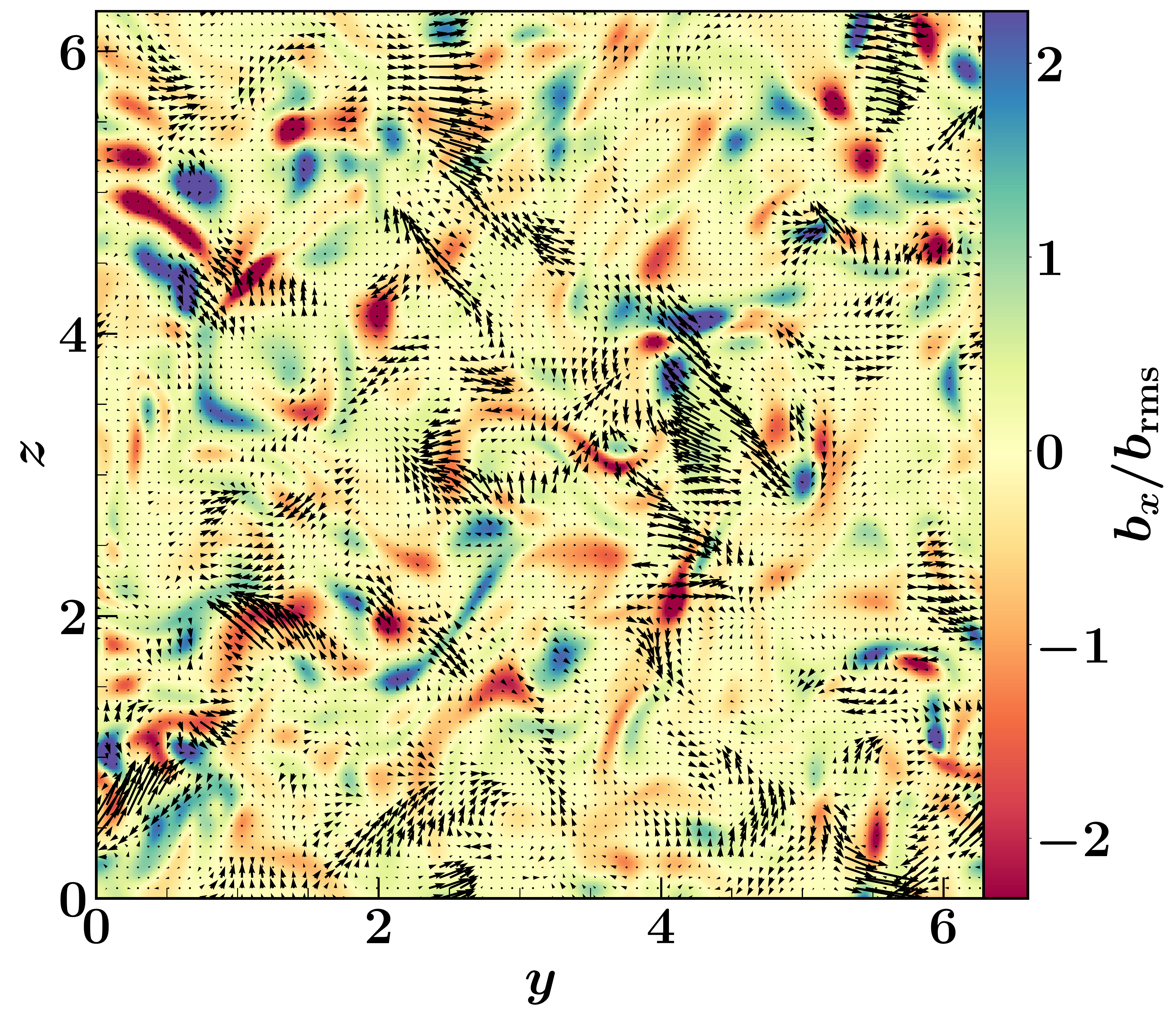}
      \includegraphics[width=\columnwidth]{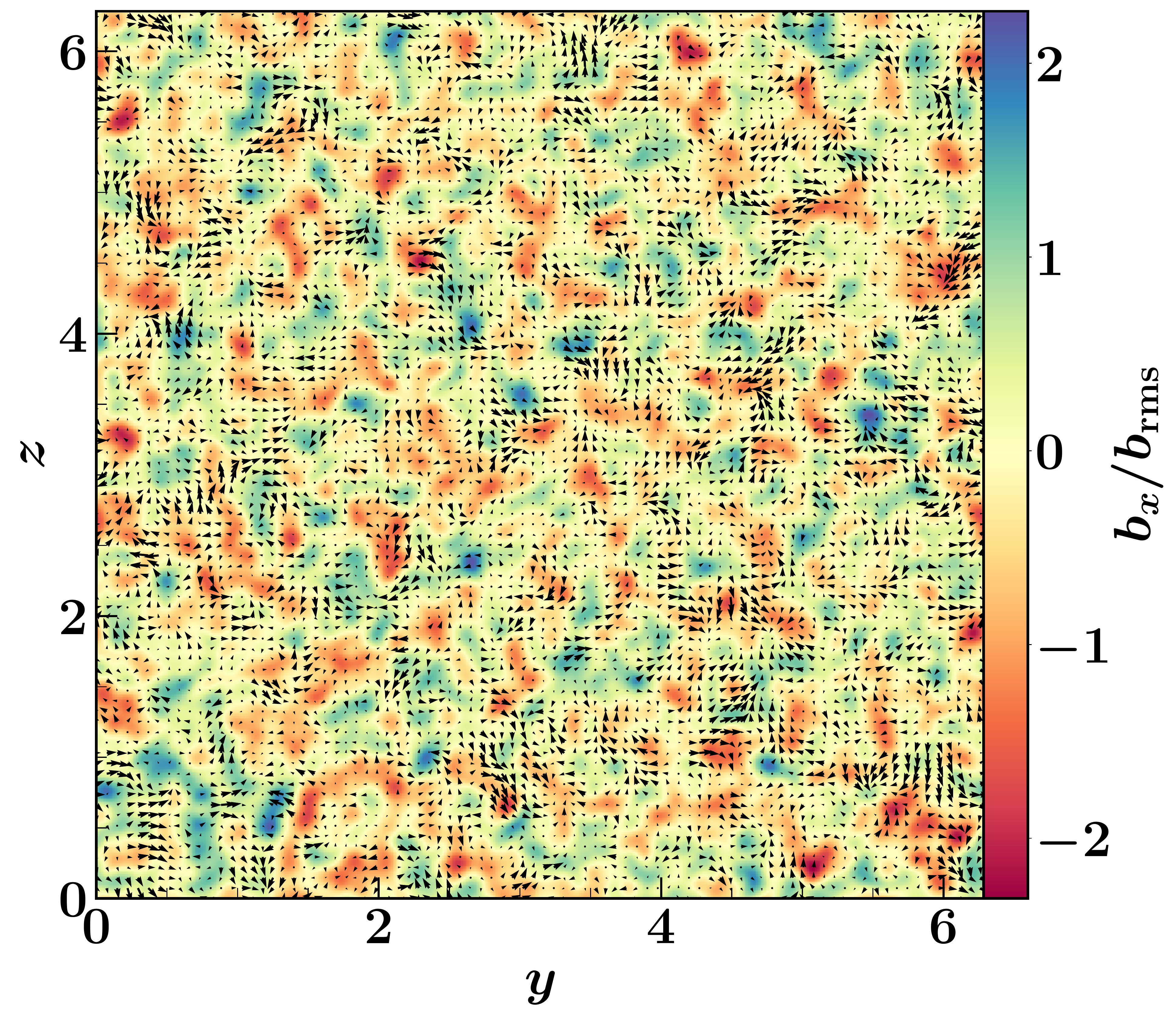}
      \caption{2D cut in the $yz$--plane through the middle of the domain with vectors for $(b_y/\brms,b_z/\brms)$ and colours showing the magnitude of $b_x/\brms$ 
               for intermittent (left) and randomized (right) magnetic fields. For intermittent magnetic field the colours are saturated for both 
               positive and negative values to match the scale of the randomized field (see the $x$-axis of \Fig{bpdf} for the actual difference in numbers).
               The intermittent magnetic field is more ordered and stronger in the filaments whereas the randomized field lacks such structures.}
      \label{bvec} 
\end{figure*} 

\begin{figure} \centering
      \includegraphics[width=\columnwidth]{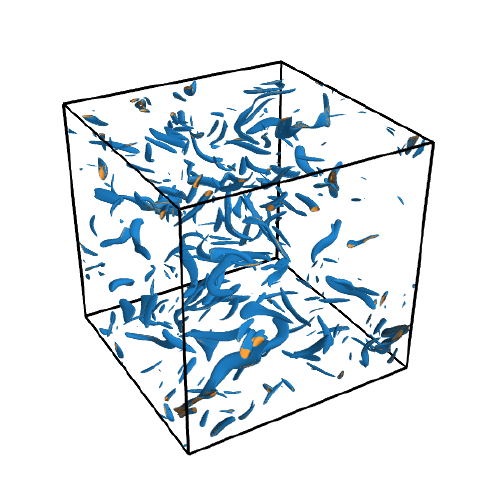}\quad
      \caption{The figure shows isosurfaces of $b^2/\brms^2 = 12$ (blue) and $b^2/\brms^2 = 15$ (yellow) for the dynamo-generated magnetic field at $\Rm = 3182$. It is intermittent,
               showing long filaments with large gaps between filaments.}
      \label{b2} 
\end{figure}
\begin{figure}
      \includegraphics[width=0.9\columnwidth]{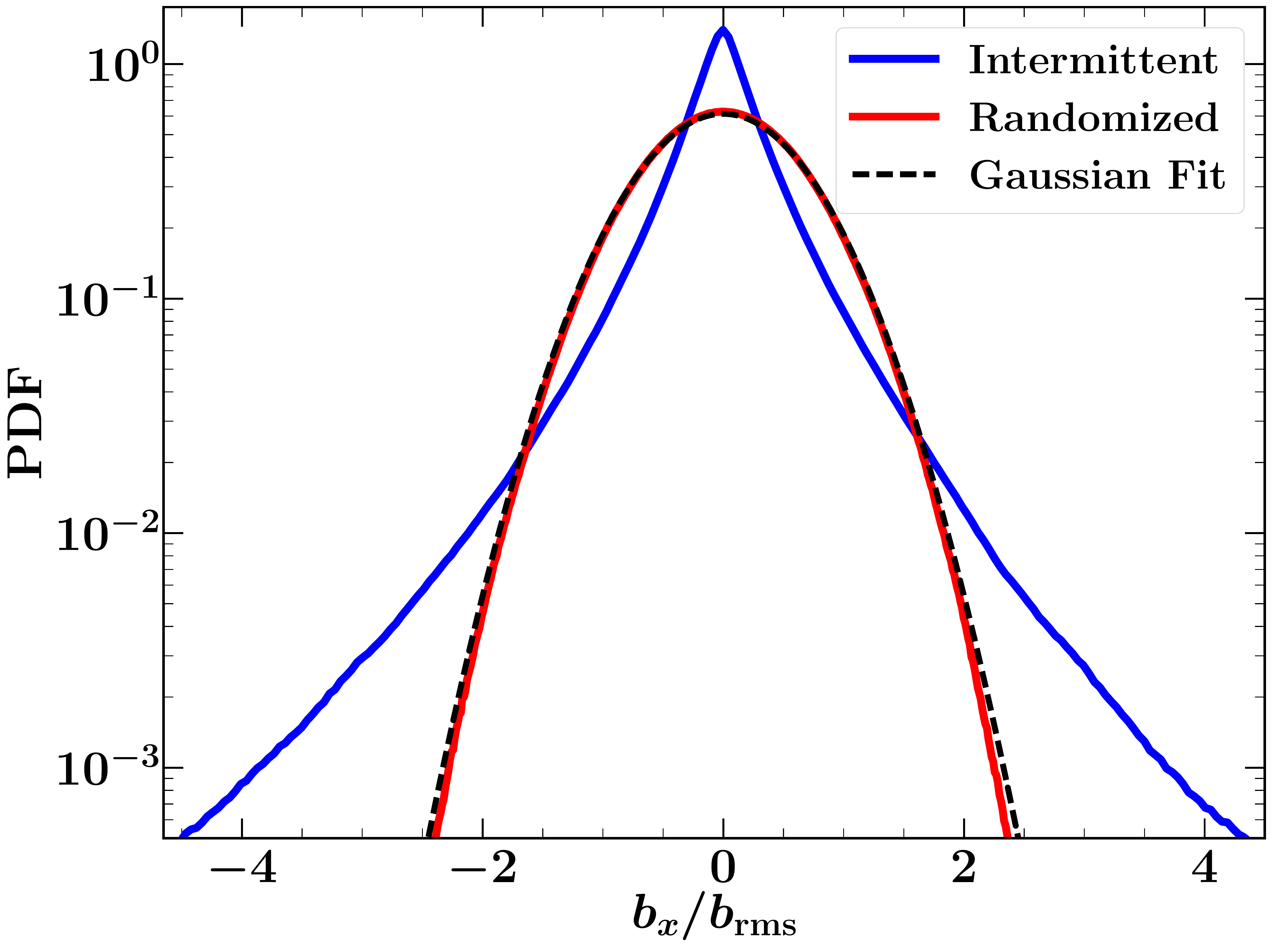}
      \caption{The figure shows PDFs of $b_x/\brms$ for intermittent (blue) and randomized (red) magnetic fields. 
               Both have zero mean but the intermittent magnetic field has long tails whereas the randomized field (obtained by Fourier phase
               randomization) has a Gaussian PDF (dashed).}
      \label{bpdf}
\end{figure}

We construct the velocity field, $\vec{u}$, by superposing Fourier modes with a range of wavenumbers, $k$,
and a chosen energy spectrum $E(k)$,
using the same prescription as \cite{FungEA1992} and \cite{Wilkin2007}:
\begin{equation} \label{ks} 
\vec{u}(\vec{x},t) = \sum_{n=0}^{N-1} \left[\vec{C}_{n}(\vec{k}_n) \cos \phi_n + \vec{D}_{n}(\vec{k}_n) \sin \phi_n\right],
\end{equation}
where $\phi_n = \vec{k}_n \cdot \vec{x} + \omega_n t$, $\vec{k}_n$ is a randomly oriented wavevector of magnitude $k_n$,
and $\omega_n=[k_n^3 E(k_n)]^{1/2}$ is
the frequency at that scale.
The vectors $\vec{C}_n(\vec{k}_n)$ and $\vec{D}_n(\vec{k}_n)$ have random directions in the 
plane perpendicular to $\vec{k}_n$,
so that the flow is incompressible ($\nabla \cdot \vec{u} = 0$).
Their magnitudes determine
the power spectrum $E(k)$,
and are chosen
such that $E(k) \propto k^{-5/3}$
and the rms velocity is $u_0$.
We chose $N=40$ with $\vec{k}_n$ such that the flow is periodic, and with distinct $k_n$ between $2 \pi/L$ to $8 \pi/L$, where $L=2\pi$ is the width of our domain which is also equal to the outer scale, $l_0$, of the velocity flow.
This flow acts as a dynamo if $\Rm > \Rmc \simeq 1000$, and produces a spatially intermittent magnetic field,
as can be seen in the left-hand panel of \Fig{bvec} and \Fig{b2}. 

The presence of structures in the magnetic field affects the propagation of cosmic rays, especially for low energy particles.
For such particles, intermittency enhances cosmic ray diffusion \citep{SSSBW17}. 
Here we are interested in exploring the correlation between the magnetic field and cosmic rays.
The intermittent nature of this dynamo-generated magnetic field provides an excellent test case for this study since there are localized regions of strong magnetic field.

We also consider a Gaussian random magnetic field having 
\textit{the same power spectrum} $M_b(k)$ as the intermittent magnetic field. 
Such a Gaussian random field is obtained as follows: First, a spatial Fourier transform of the intermittent magnetic field is taken and then each complex mode is multiplied with a random phase. 
Then taking an inverse Fourier transform gives a Gaussian randomized magnetic field with unchanged $M_b(k)$, but where coherent structures have been destroyed \citep[Chapter~7 in][]{Biskamp2003,Snodin2013,SSSBW17}.
The structural difference between intermittent and Gaussian random magnetic field is illustrated in \Fig{bvec} which 
shows the magnetic fields in a 2D cut through the middle of the numerical domain, with colours showing the third component.
Figure~\ref{b2} shows the filamentary structure of the 
intermittent magnetic field (at $b^2/\brms^2$ of order ten).
The probability distribution function (PDF) of a single component of both the intermittent and randomized field is shown in \Fig{bpdf}. The intermittent field has long heavy tails, whereas the
randomized field has a Gaussian probability distribution. 
\subsection{Large-scale magnetic field} \label{mean}
The ISM contains both fluctuating (small-scale)
and mean (large-scale) magnetic fields 
\citep[Chapter~5 in][]{KleinFletcher2015,Beck1996,Beck2016}. 
The large-scale component is correlated over several $\kpc$ whereas the small-scale 
component has a correlation length less than the correlation scale of turbulence ($\lesssim 0.1 \kpc$).
The small-scale and the large-scale components of the magnetic field contain comparable energies.
By comparing large-scale magnetic field models (with resolution of $\sim 75\pc$) of a spiral galaxy with observations, 
\cite{Moss2007} found that a spatially uniform distribution of cosmic rays 
was better at matching the observations than an equipartition assumption between cosmic rays and large-scale  
magnetic field. This suggest that the cosmic rays may not be correlated with large-scale magnetic field.

To explore the effects of a large-scale magnetic field on the propagation of cosmic rays,
we add a uniform mean field directed along the $x$-axis
to the random magnetic field as described in Section~\ref{MFGRF}.
We consider several values for
the ratio of the mean field $B_0$ to the random field, $\brms$,
up to $B_0/\brms = 3$.

\begin{figure*} \centering 
	\includegraphics[width=\columnwidth]{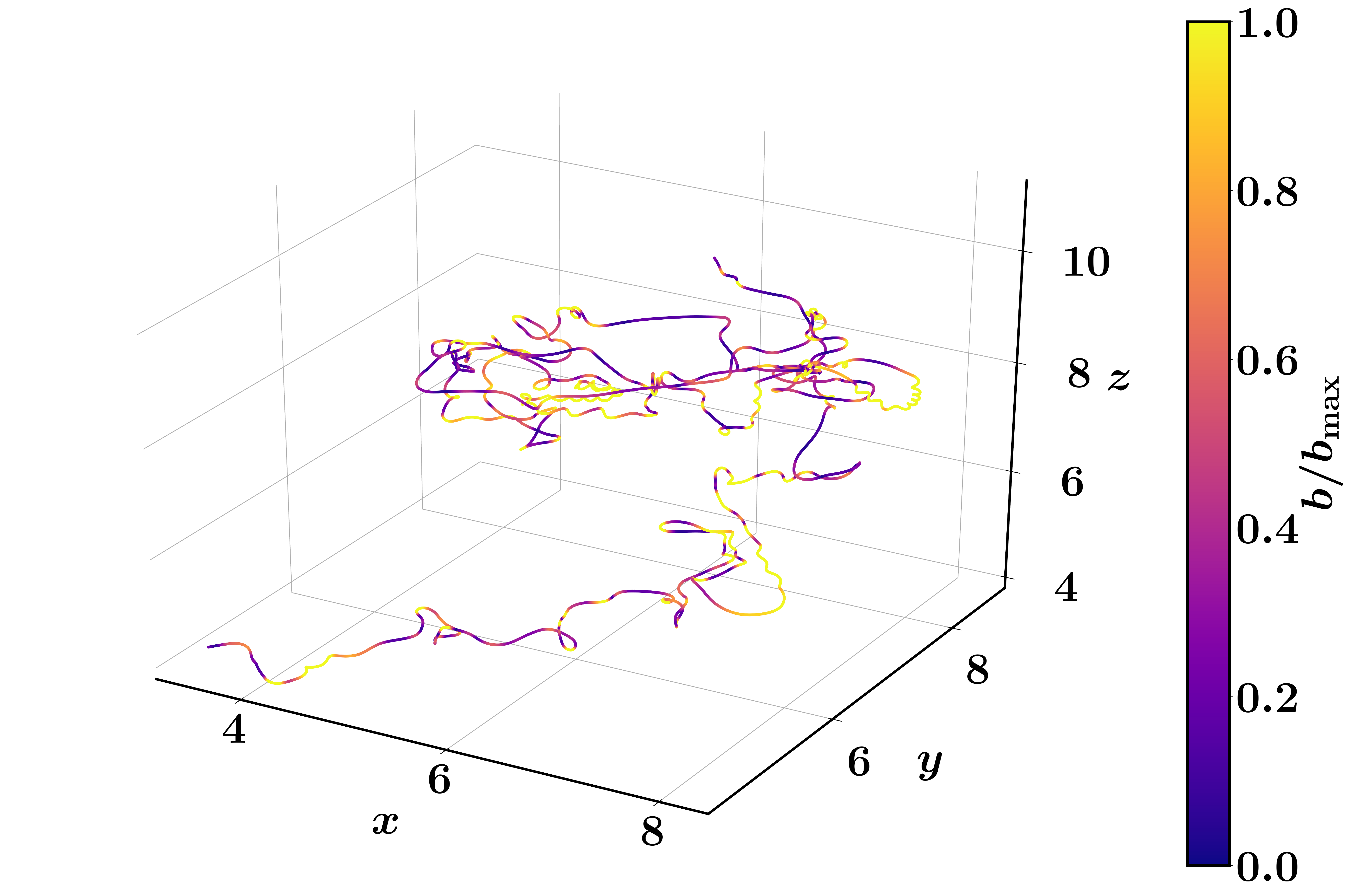}
	\includegraphics[width=\columnwidth]{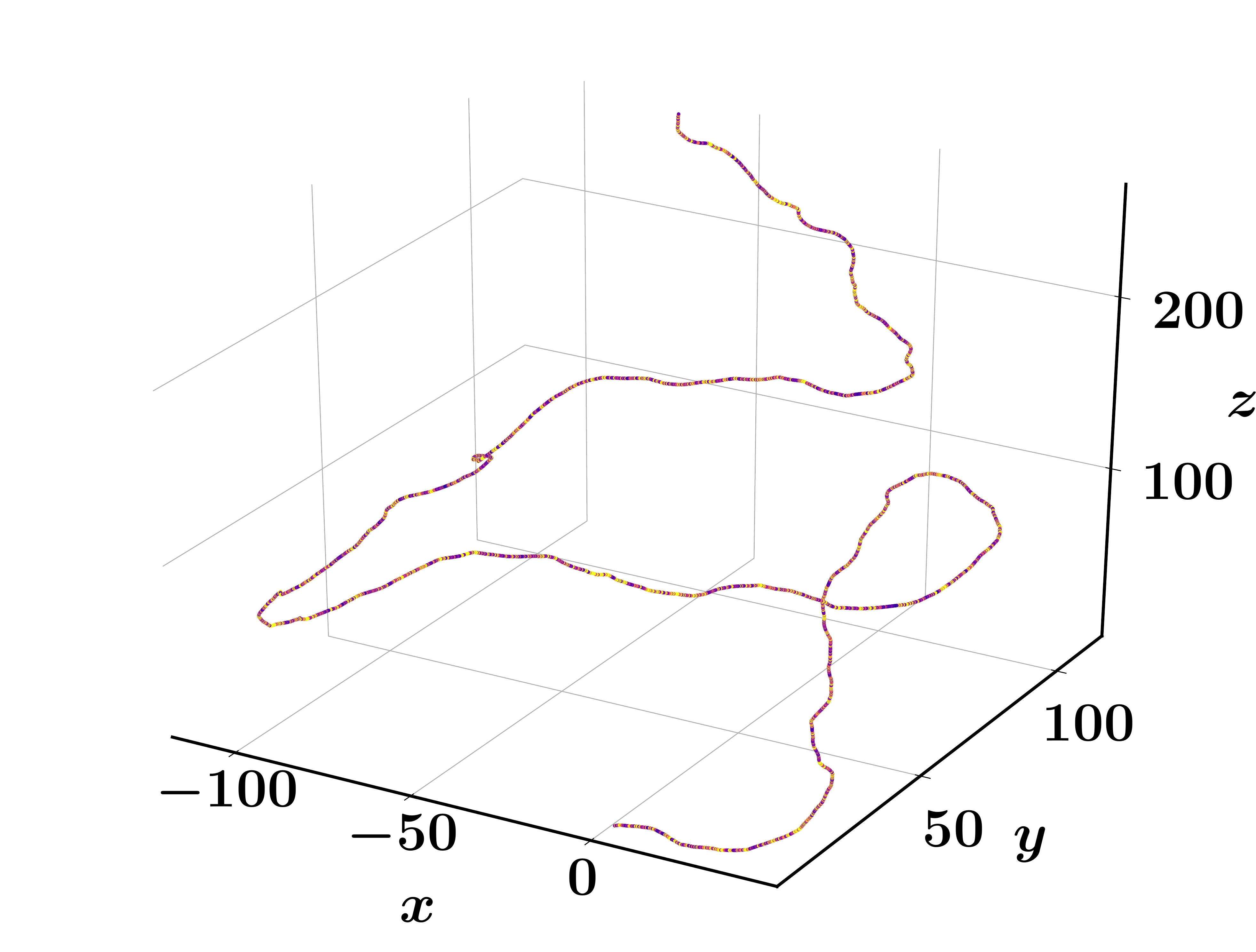}
	\caption{Single particle trajectories for $\Lar/l_0 = 0.011$ (left-hand panel) and $0.318$ (right-hand panel) in 
		the intermittent magnetic field of the left-hand panel of Fig.~\ref{bvec} and Fig.~\ref{b2}. Colours shows the
                strength of the magnetic field along the trajectory normalized
                to its maximum value along the part of trajectory shown.  
		For $\Lar/l_0 = 0.011$, the particle path is more tangled than for $\Lar/l_0 = 0.318$
		where the particle motion is almost ballistic between rare scattering events.}
        \label{traj}
\end{figure*}

\section{Cosmic ray propagation} \label{crpropg}
The propagation of cosmic rays in random magnetic fields is largely determined by the ratio between the Larmor radius $\Lar$ of 
the particle gyration and the length scale $l_b$ of magnetic field variations, e.g., its correlation length.
The Larmor radius and Larmor frequency of a relativistic particle of 
rest mass $m$ and charge $q$, 
travelling at a speed $v$ in a magnetic field of strength $B$, are given by
\begin{equation} \label{larmor_radius}
 \Lar = \frac{\gamma m c v}{qB} \qquad \mbox{and} \qquad \omega_0 = \frac{v}{\Lar},
\end{equation}
respectively,
where $\gamma = \left(1 - v^2/c^2\right)^{-1/2}$ 
is the Lorentz factor, and $c$ is the speed of light in a vacuum. 
The correlation length $l_b$ of an isotropic random field
with power spectrum $M_b(k)$ is defined as \citep{Monin_Yaglom1971} 
\begin{equation}\label{l_b}
l_b = \frac{\pi}{2} \frac{\int_0^{\infty} k^{-1} M_b(k)\, \dd k}{\int_0^{\infty} M_b(k) \,\dd k} .
\end{equation}
The correlation length of a magnetic field produced by the fluctuation dynamo action is significantly
smaller than $l_0$, at least in the kinematic stage 
where $l_b/l_0$ is of order $\Rm^{-1/2}$ \citep{Kaz68,ZRS90,SCTMM04,BS2005}. 
For $\Rm=3182$, we have $l_b/l_0\approx0.0244$.

\begin{figure*}  \centering
	\includegraphics[width=1.8\columnwidth]{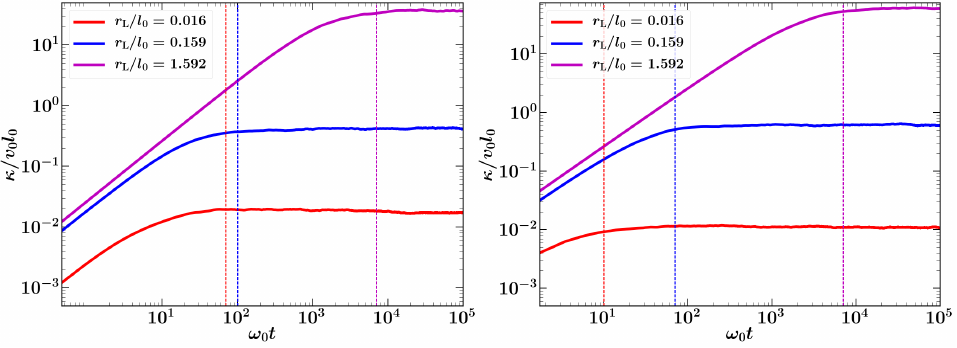}
	\caption{Normalized cosmic ray diffusivity $\kappa/v_0 l_0$ as a function of normalized time $\omega_{0} t$ for $\Lar/l_0 = 0.016,0.159,1.592$ for the intermittent (left) and randomized (right) magnetic fields 
         shown in Fig.~\ref{bvec}.
		Here  $\omega_0 = v_0/\Lar$ is the Larmor frequency based on $\Brms$.
		The dashed lines of the corresponding colours show the time $\tdiff$ after which the propagation becomes diffusive:
		$\kappa(t)\approx\kappa$ at $t>\tdiff$. For low energy particles ($\Lar \le l_b$), the diffusivity in the intermittent magnetic field is larger than in the randomized
		(Gaussian)  field of identical power spectrum.}
	\label{tdiff}
\end{figure*}

\subsection{Test particle simulations of cosmic rays}\label{TPSCR}
 
We consider relativistic charged particles propagating in a 
static magnetic field, $\vec{B}(\vec{x})$.
The trajectory of each particle satisfies

\begin{equation} \label{lf} 
\oderiv{^2\vec{r}}{t^2} = \frac{v_0}{\Lar}\,\oderiv{\vec{r}}{t} 
       \times \frac{\vec{B}}{\Brms}, 
\end{equation}
where $\vec{r}$ is the particle's position,  $v_0$ is its speed,
$\vec{B}/\Brms$ is the total magnetic field normalized to its rms value $\Brms$,
and $\Lar$ is the Larmor radius (defined with respect to $\Brms$).
The time-scale over which interstellar magnetic fields change significantly is of the order
of the eddy turnover time in interstellar turbulence, $10^7 \yr$ \citep{Beck1996}.
This time-scale is longer than the (diffusive) confinement time of cosmic rays in galaxies, which is of the order of $10^6 \yr$
\citep{BBDP90}.
It is therefore customary to neglect any time dependence of the magnetic field in \Eq{lf} and, 
correspondingly, neglect any electric fields \citep[e.g.,][]{GJ1999,Casse_et_al2002}.
This means that the speed of each particle remains constant. This is equivalent
to neglecting particle acceleration, a process physically distinct from diffusive particle 
propagation (which is the focus of this article) which is due to scattering of
cosmic ray particles by magnetic inhomogeneities. Since we consider static magnetic field, the diffusive 
reacceleration \citep{Strong2007,GBS15},
which is due to moving magnetic inhomogeneities, is also neglected.

We solve \Eq{lf} numerically for an ensemble of cosmic ray particles (specifically, $8192$),
all of the same speed $v_0$, but giving each a random initial position and velocity direction.
The initial conditions are chosen randomly, and 
are uniformly distributed over all positions and directions, but do not very uniformly fill the magnetic field cube,
due to there only being a finite number of particles.
For a given energy of particle, we find the smallest Larmor time ($2\pi/\omega_0$) based on the 
maximum magnetic field in the domain and then fix the time step as $0.001$ times the
smallest Larmor time, to ensure that we carefully resolve all particle gyrations. We also check that the energy 
is conserved (as far as is permitted by the numerical scheme) throughout the total propagation time $T$.
For a given magnetic field configuration, the 
nature of the trajectories depends only on the parameter $\Lar/l_0$, which is indicative of the particle energy, as illustrated in \Fig{traj}.
By construction, the static magnetic field through which the particle propagates is periodic in all three
directions, with period $L= 2 \pi$.
Even though the magnetic field is periodic, 
the particle trajectories are not: they enter and leave the domain at different points. 
There is an important distinction between the Eulerian frame of the computational domain and the Lagrangian frame moving with each particle. 
Whilst the magnetic field is periodic in the Eulerian sense, there is no periodicity in the magnetic field {\it along each particle trajectory} (however many times the particle enters and leaves the domain).

A high-energy particle, with $\Lar \gg l_b$, is typically deflected by only a small angle of order $l_b/\Lar$
over a distance $l_b$ and its trajectory is, therefore, rather insensitive to the structural properties of the 
magnetic field at scales smaller than $\Lar$. 
By the central limit theorem, the statistical properties of an ensemble of
cosmic ray particles become Gaussian after a large number of such deflections.
Particles of smaller energies, $\Lar \lesssim l_b$,
are more sensitive to the fine structure of the magnetic field,
and it is not obvious how the spatial distribution of such cosmic
rays will be related to the magnetic energy density, especially given that their
diffusion tensor is sensitive to magnetic intermittency  \citep{SSSBW17}.
In particular,
the distribution of cosmic rays may be intermittent in a spatially intermittent magnetic field.
Figure~\ref{traj} shows the trajectories of particles of low and high energy (left- and right-hand panels, 
respectively). The latter move
faster (note the different axis scales in the two panels) and, at the scale of the left-hand panel,
the trajectory of the higher-energy particle is nearly straight almost everywhere.

The random nature of the magnetic field makes the particle propagation diffusive at sufficiently
large spatial and temporal scales. Without a mean field, the propagation is isotropic.
We therefore calculate the isotropic diffusion coefficient as the limit of the
finite-time diffusivity $\kappa(t)$,
\begin{equation}  \label{diff}
\kappa = \lim_{t \to \infty} \kappa(t),
\qquad
\kappa (t) = \frac{1}{6t} \langle |\vec{r}(t)-\vec{r}(0)|^2 \rangle,
\end{equation}
where
the angular brackets denote averaging over the ensemble of particles.
Figure~\ref{tdiff} shows $\kappa (t)$ for the two magnetic fields shown in Fig.~\ref{bvec},
one intermittent and the other statistically Gaussian, and for several values of $\Lar/l_0$.
In each case there is an initial phase of ballistic particle motion, in which $\kappa(t)$ is 
approximately linear in $t$, followed by a diffusive phase where $\kappa(t)$ 
settles to its asymptotic value. The start of the diffusive phase, $\tdiff$,
is the time when the slope of $\kappa(t)$ becomes small, $\dd\kappa/\dd(\omega_0 t) \simeq 10^{-6}$;
this time is indicated by a vertical dashed line in \Fig{tdiff}.

To obtain the number density of cosmic rays, $\ncr$, from the test particle simulations
we calculate the coordinates of each particle modulo $L=2\pi$,
i.e., relative to the periodic magnetic field. Next,
we divide the periodic domain into $512^3$ cubes and
count the number of particles within each cube.
The size of the cubes was chosen to match the spatial resolution of the magnetic field,
which was obtained from a dynamo simulation on a $512^3$ grid,
but we have checked that the results are not very sensitive to the exact size of the cubes.
The result is the instantaneous number density of the particles $\tilde{n}(\vec{x},t)$.
We then average the density of particles within each cube over a sufficiently long
period $T$ ($\gg\tdiff$), to obtain the cosmic ray density 
$\ncr (\vec{x})=(T- \tdiff)^{-1}\int_{\tdiff}^T \tilde{n}(\vec{x},t')\,\dd t'$. 
We have checked that the results are not dependent on the sampling time ($\dd t'$)
as long as it is smaller than a few times the Larmor time ($2\pi/\omega_0$).
But for lower sampling rate the simulation has to be averaged over a longer total time (higher $T$) to collect sufficient statistics. 
We note that different energies were simulated over different periods $T$ to obtain
roughly the same $\langle {\ncr(\vec{x})} \rangle$ for all energies.

 \begin{figure*} \centering
 	\includegraphics[width=0.9\columnwidth]{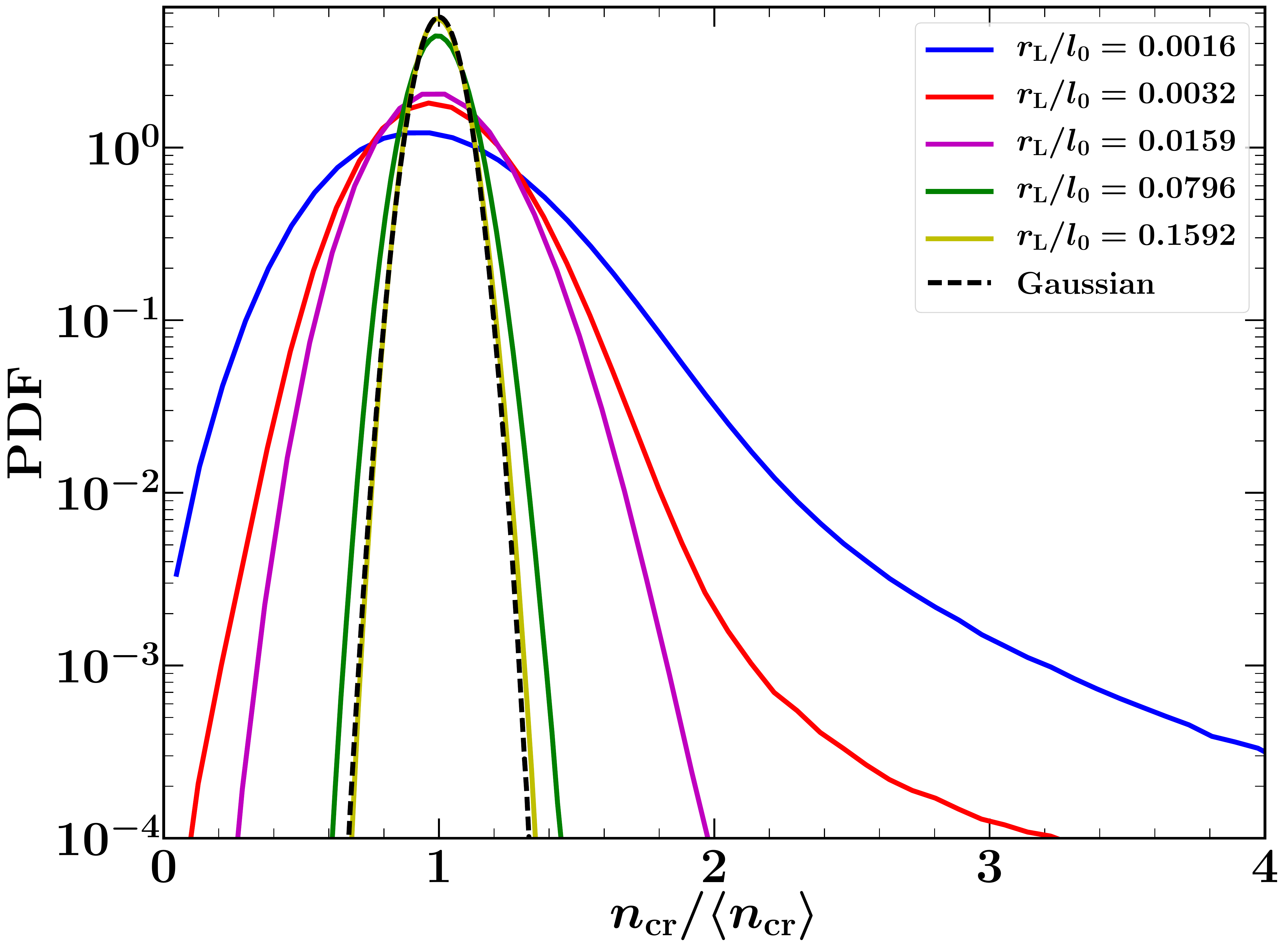}
 	\includegraphics[width=0.9\columnwidth]{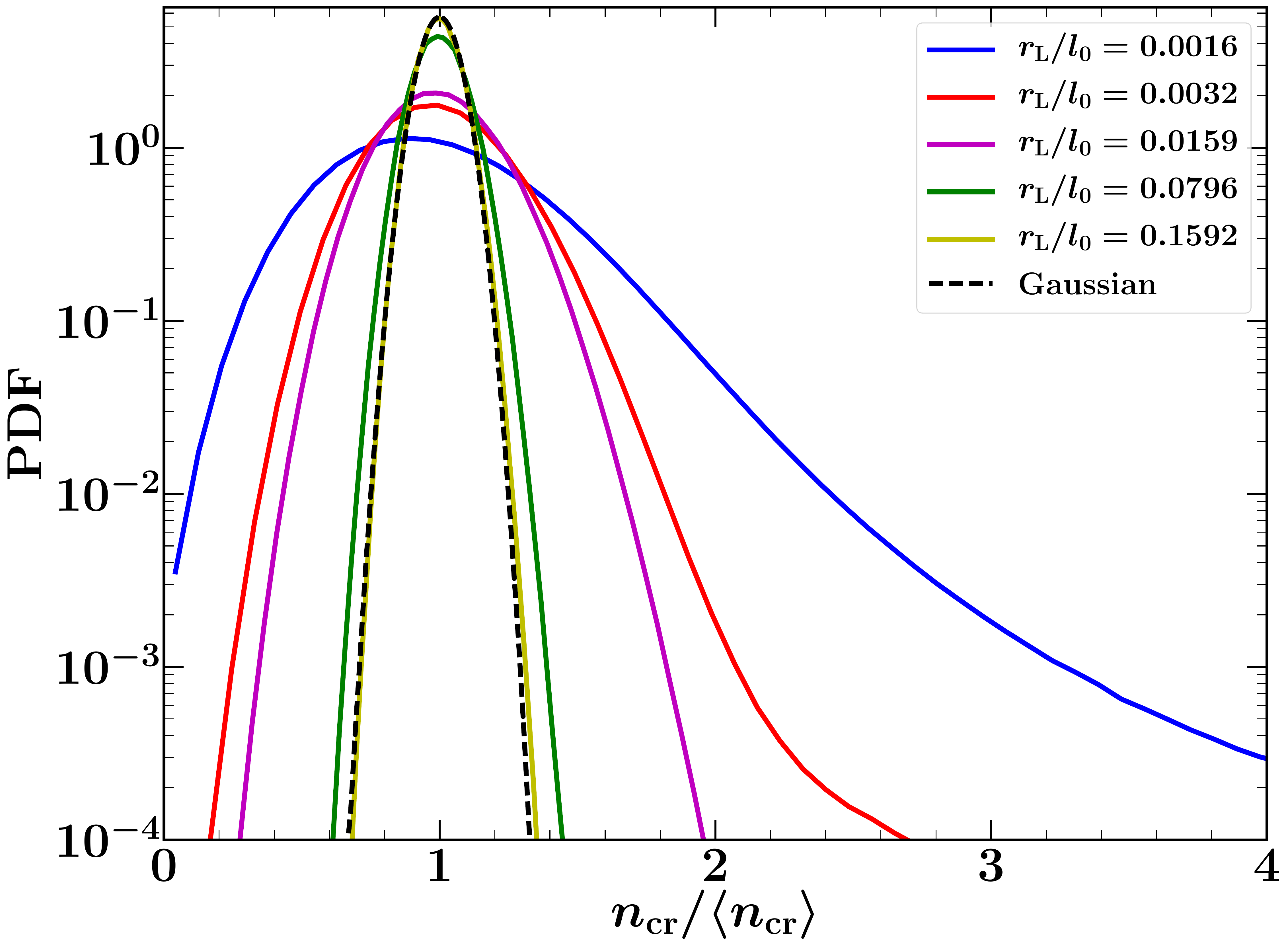}
 	\caption{Probability density function (PDF) of the relative number density of cosmic ray particles, $\ncr/\ncravg$, for various $\Lar/l_0$ in intermittent (left panel) and randomized (Gaussian) magnetic fields (right panel), with no mean-field and no pitch angle scattering (PAS). 
 		Long tails are a signature of intermittent structures in the cosmic ray distribution. 
 		For high-energy particles, the distribution is nearly Gaussian (with width increasing as energy decreases) in both intermittent and Gaussian magnetic fields, but
 		below a certain energy ($\Lar \lesssim l_b$) long tails develop.  A black dashed line shows the 
 		PDF of a random variable drawn from a Gaussian distribution with unit mean value and standard deviation of $0.07$. }
 	\label{ncrpdf1}
 \end{figure*}
 
 \begin{figure*} \centering
  	\includegraphics[width=0.9\columnwidth]{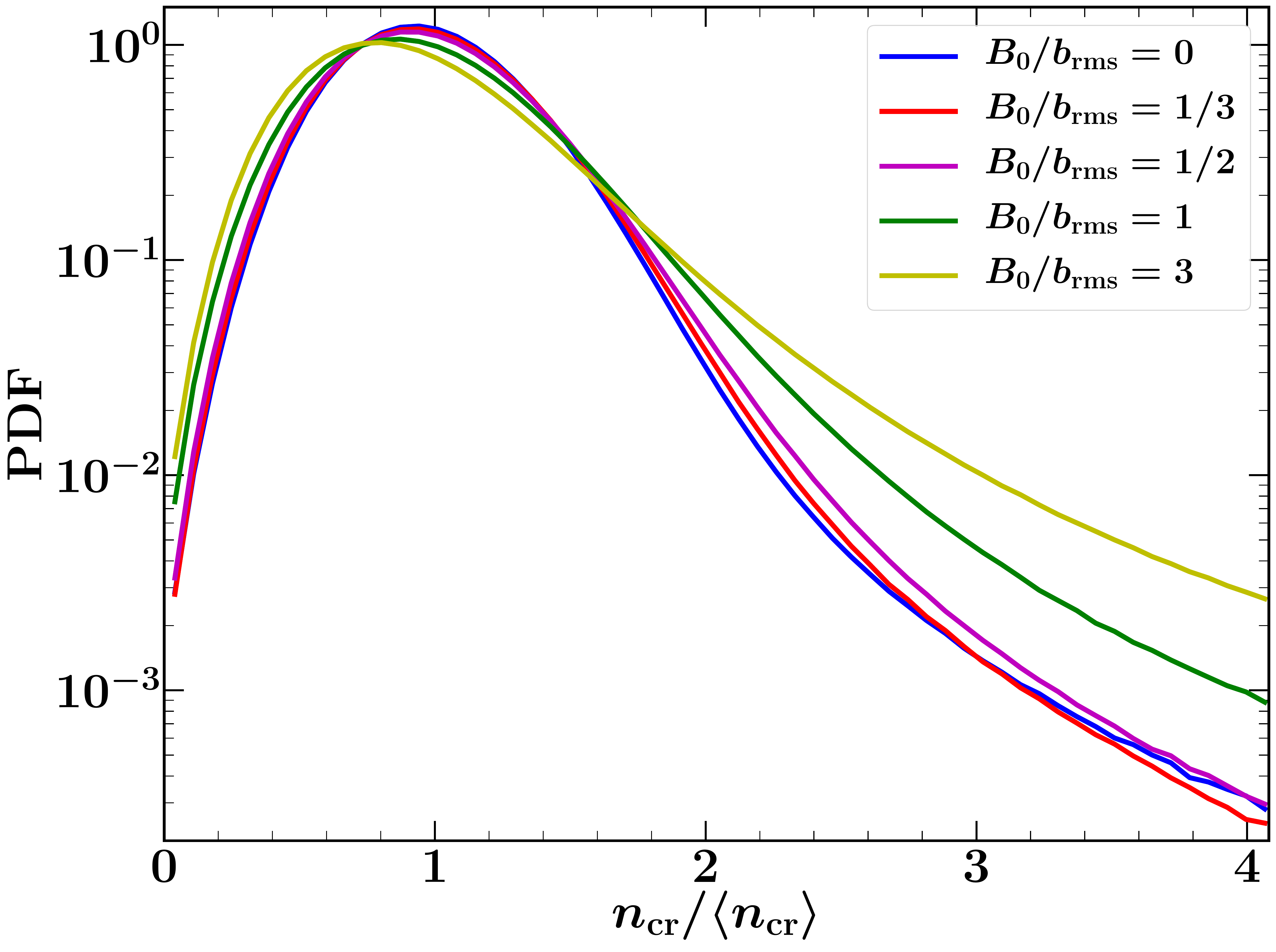}\qquad
 	\includegraphics[width=0.9\columnwidth]{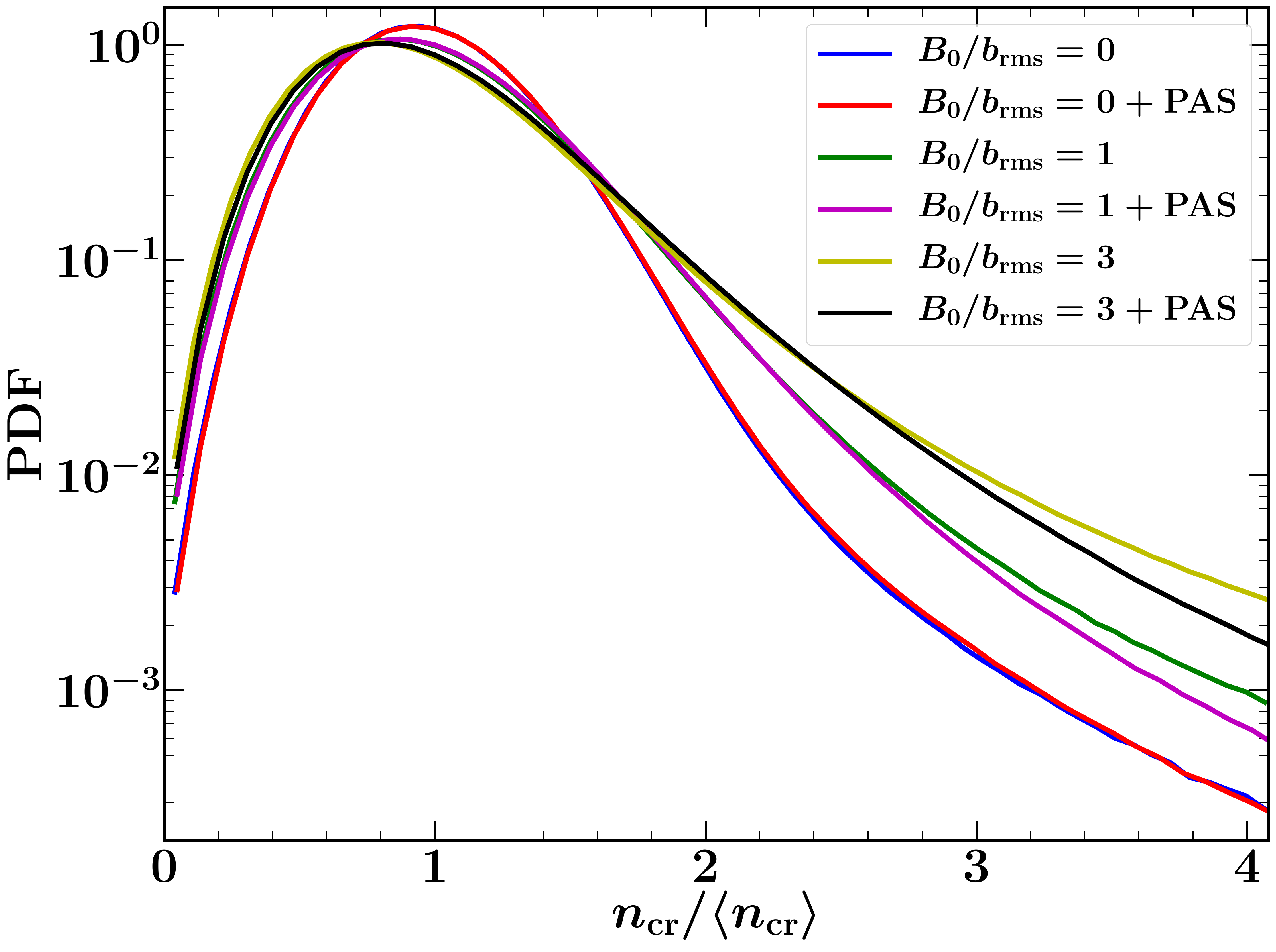}
 	\caption{PDF of $\ncr/\ncravg$ for $\Lar/l_0 = 0.0016$ in the intermittent field of  Fig.~\ref{b2} with an imposed mean field of various magnitudes $B_0$ (left-hand panel) 
                 and with the pitch angle scattering further included (right-hand panel). Intermittency in the cosmic ray distribution increases as the mean field becomes stronger, especially for $B_0/\brms \ge 1$,
                 manifested in heavier tails at larger $\ncr/\ncravg$. The pitch angle scattering enhances diffusion and thus decreases the level of intermittency.}
 	\label{ncrpdf2}
 \end{figure*}

\subsection{Magnetic field at the Larmor scale} \label{pas}
Cosmic ray particles are especially sensitive to magnetic fluctuations at a scale 
comparable to their Larmor radius. As described by \cite{Kulsrud2005},
when a cosmic ray particle encounters such a magnetic fluctuation,
its pitch angle $\theta$ defined via
\begin{equation} \label{mu}
	\mu \equiv \cos\theta = \frac{\dsp \vec{v} \cdot \vec{B}}{\dsp |\vec{v}| |\vec{B}|}\,,
	\qquad \vec{v}=\oderiv{\vec{r}}{t}\,,
\end{equation}
changes by
\begin{equation}\label{deltatheta}
	\delta \theta = -\pi \frac{\delta B}{B} \cos\chi\,,
\end{equation}
where
$\delta B/B$
is the ratio of the fluctuation amplitude
at the Larmor scale to the local mean magnetic field
(i.e., the field averaged over a scale significantly larger than $\Lar$), and
$\chi$ is the relative phase between the cosmic ray velocity vector and the wavevector
of the magnetic fluctuation with which the particle interacts.

Magnetic fluctuations at the Larmor scale can be 
a part of a magnetic energy spectrum that extends from larger scales or be
excited by cosmic rays themselves via the streaming instability \citep{KP1969,Wentzel1974,Kulsrud2005}.
The latter are usually referred to as `self-generated waves'.
The scattering due to magnetic fluctuations at the Larmor scale is referred to as pitch angle scattering, and \citet{DZ2014} stress its 
importance for cosmic ray diffusion. The magnitude of the magnetic fluctuations that are associated with
pitch angle scattering due to self-generated waves is estimated in Appendix~\ref{fluc} as $\delta B/B\simeq10^{-2}$ in the hot
interstellar gas.

The spectrum of hydromagnetic turbulence in the ISM extends to very small 
scales. \citet{SCDHHQT09} suggest that the spectrum of kinetic Alfv\'en waves is truncated by
dissipation at scales as small as the 
thermal electron Larmor radius, which is approximately $3\times10^6\cm$ in the warm ionized ISM.
This scale is much smaller than the Larmor radius of the relativistic particles of cosmic rays.
However the spectrum of Alfv\'en wave turbulence is rather steep:
$k^{-5/3}$ at larger scales \citep[above $1\pc$ --][]{BS2005} where the gas is collisional and,
at smaller scales, $k_\parallel^{-2}$ 
for the perturbations parallel to the magnetic field
\citep[which matter for the cosmic ray scattering --][]{FG2004}.
For such steep spectra, the relative magnitude of the magnetic fluctuations at
the Larmor radius of a $5\GeV$ particle is of the order of $\delta B/B\simeq10^{-4}$, which is negligible
in comparison with the self-generated waves.

Within our model the magnetic field is imposed and the streaming instability cannot occur. We
therefore parametrize the cosmic ray scattering by self-generated waves. This is done 
by rotating the velocity vector of each particle every Larmor time ($2 \pi/ \omega_0$) 
by an angle given by \Eq{deltatheta}, with $\delta B/B = 10^{-2}$ and $\chi$ uniformly distributed between $0$ and $2\pi$.
Throughout the text, this is referred to as pitch angle scattering (PAS).

\begin{figure*} \centering 
	\includegraphics[width=1.95\columnwidth]{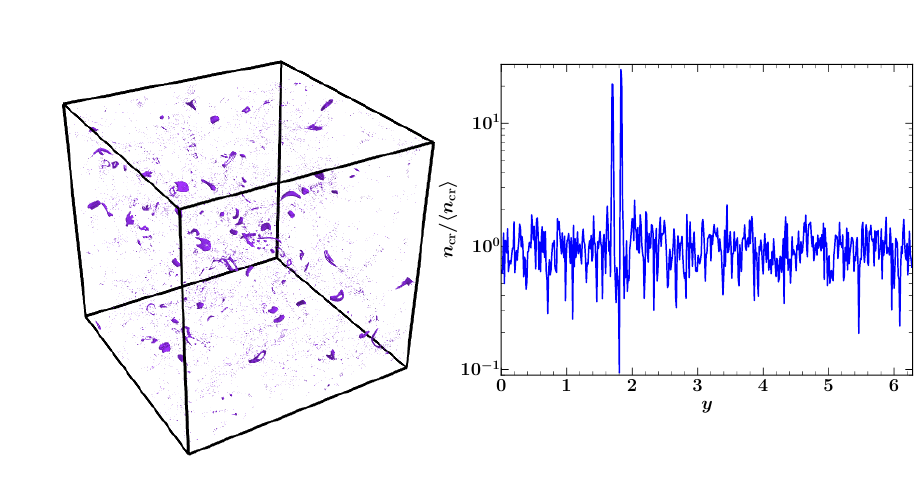}
	\caption{Isosurfaces of the number density of cosmic rays at $\ncr/\ncravg = 3.5$  for $\Lar/l_0 = 0.0016$ in an 
		intermittent magnetic field showing a highly inhomogeneous distribution (left-hand panel). 
		The right-hand panel shows the variation of the relative number density of the particles along the straight line $(x,z) = (\pi, 3.97)$, characterized by rare, 
		strong maxima against a weakly fluctuating background.}
	\label{ncrconcone}
\end{figure*}

\begin{figure*} 
	\includegraphics[width=1.8\columnwidth]{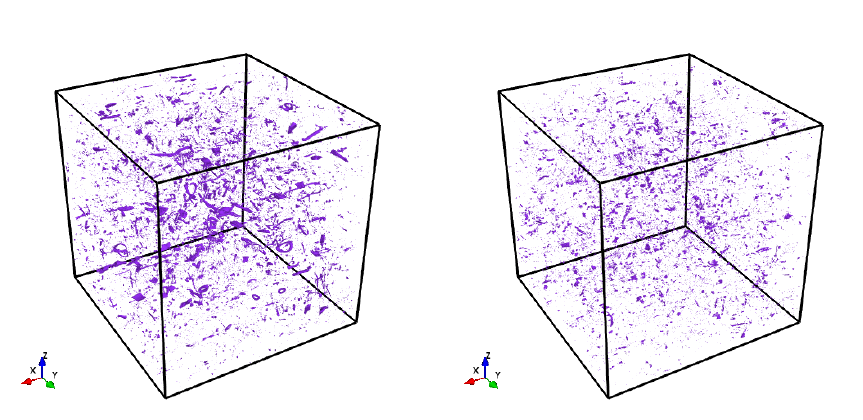}
	\caption{Isosurfaces of the number density of cosmic rays at 
		$\ncr/\ncravg = 3.5$ for $\Lar/l_0 = 0.0016$ in an intermittent magnetic field 
		with an imposed mean field of a strength $B_0/\brms = 1$ aligned with the
		$x$-axis (left-hand panel) and with pitch angle scattering further 
		added (right-hand panel).}
	\label{ncrconcsec}
\end{figure*}

 \begin{figure} \centering
 	\includegraphics[width=0.9\columnwidth]{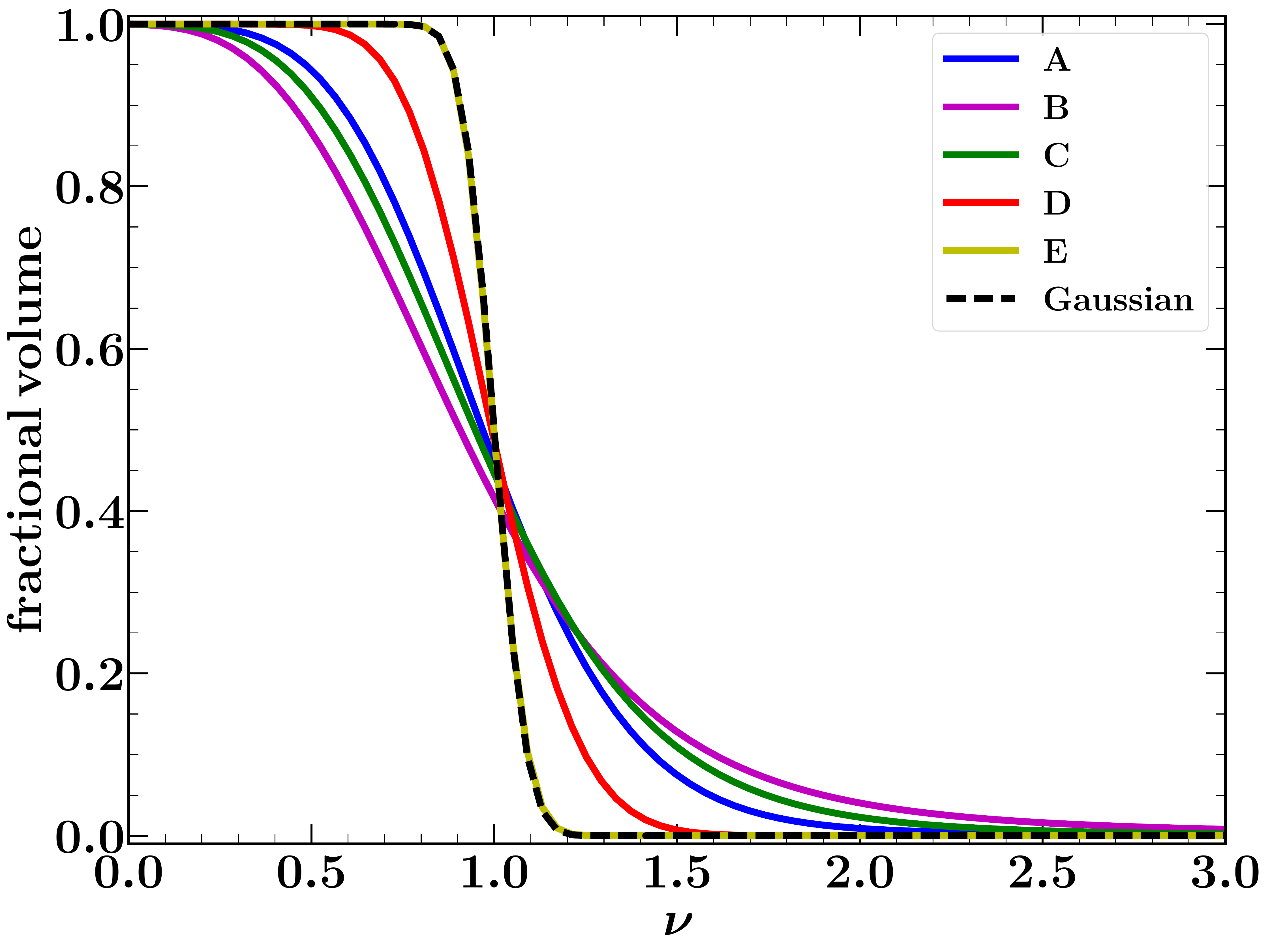}
 	\caption{The fractional volume of cosmic ray structures with 
 		$\ncr/\ncravg \ge \nu$ 
 		for Models A--E of Table~\ref{table}. A black dashed line shows the 
 		fractional volume for a random variable drawn from a Gaussian distribution with unit mean value and standard deviation of $0.07$.}
  	\label{ff}
 \end{figure}
\begin{table}
	\centering
	\caption{Representative selection of simulations, summarising the effects of particle energy, magnetic field structure and pitch angle scattering (PAS). All quantities are defined in the text.}
	\label{table}
	\begin{tabular}{cccccc} 
		\hline
		Model & $\Lar/l_0$ & $\vec{b}$ & $B_0/\brms$ & PAS & $f$ \\
		\hline
        A & $0.0016$ & Intermittent 	& 0 & no & 0.81 \\
        B & $0.0016$ & Intermittent 	& 1 & no & 0.68 \\
        C & $0.0016$ & Intermittent 	& 1 & yes & 0.84 \\
        D & $0.0016$ & Randomized 		& 0 & no & 0.96 \\
        E & $0.1592$ & Intermittent 	& 0 & no & 0.99 \\
		\hline
	\end{tabular}
\end{table}

\begin{figure} 
	\includegraphics[width=\columnwidth]{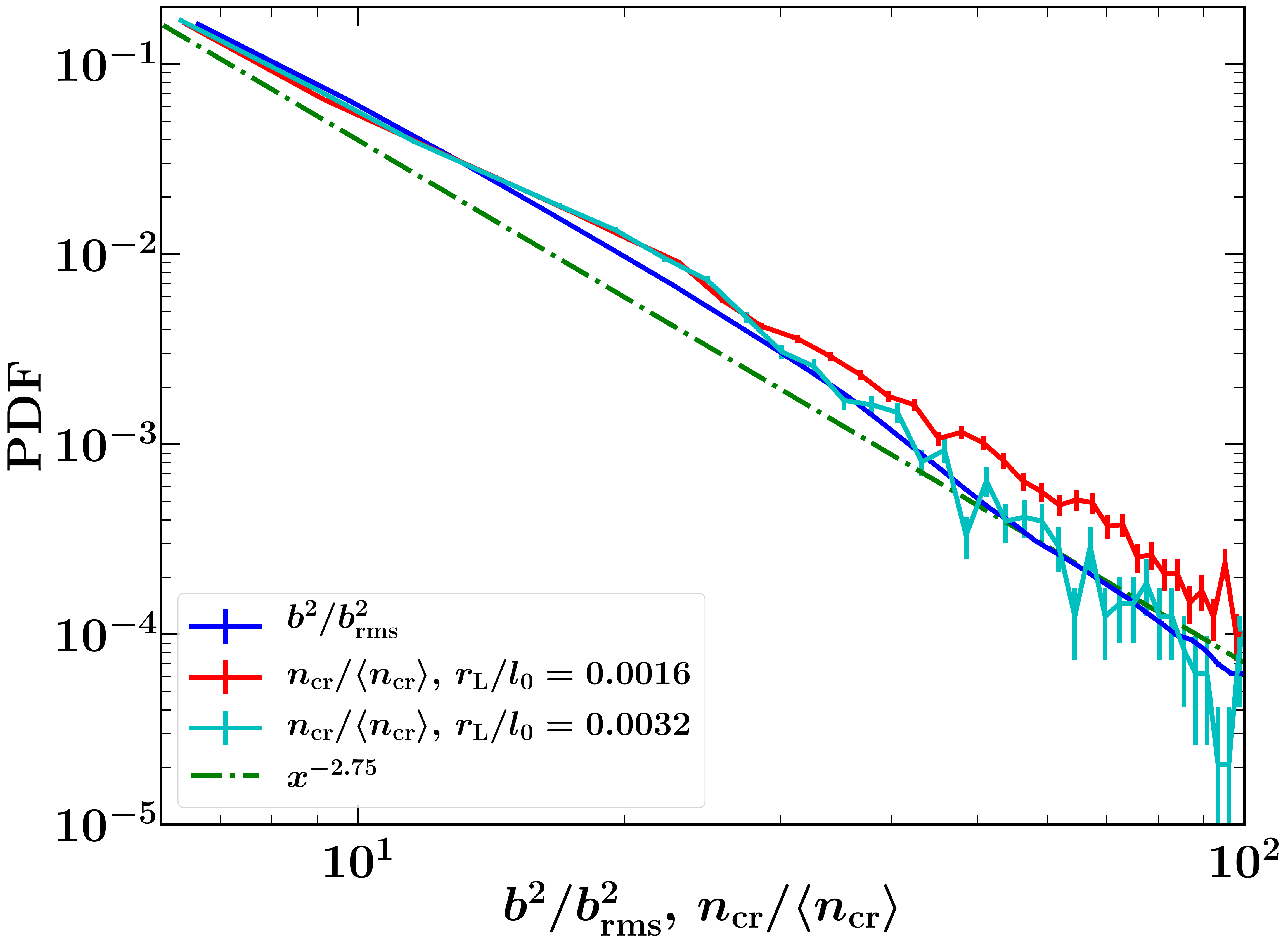}
	\caption{PDF of intermittent magnetic field energy density normalized to its rms value, $b^2/\brms^2$, and the number density of cosmic rays normalized to its mean, 
		$\ncr/\ncravg$, for $\Lar/l_0 = 0.0016, 0.0032$. 
		All three of them have power-law tails and the same exponent. Statistical errors are considerable at probability 
		densities below about $\le 5 \times 10^{-5}$.}
	\label{b2ncrpdf}
\end{figure}

\begin{figure} \centering
	\includegraphics[width=\columnwidth]{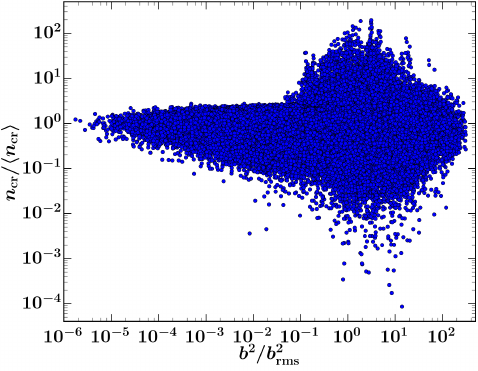}
	\caption{The scatter plot of cosmic ray number density, 
		$\ncr/\ncravg$ and magnetic energy density, $b^2/\brms^2$, in 
		intermittent magnetic field for $\Lar/l_0 = 0.0016$.}
	\label{corr2d}
\end{figure}

\section{Results}\label{results}
Figure~\ref{ncrpdf1} shows the probability density function (PDF) of the particle 
number density $\ncr$
obtained in intermittent and randomized (Gaussian) magnetic fields.
For cosmic rays of relatively high energy ($\Lar/l_b > 1$ or $\Lar/l_0 > 0.0244$),
the number density $\ncr$ is very nearly uniform in space, and its PDF is Gaussian.
At lower energies, the PDF has a long, heavy tail,
which signifies the presence of spatially localized structures in the cosmic ray
distribution. It is remarkable that the distribution of cosmic rays 
is intermittent in both intermittent and Gaussian magnetic fields. 
The PDF of $\ncr$ on including mean magnetic fields
of various strength, with (right-hand panel) and without (left-hand panel) 
particle pitch angle scattering is shown in \Fig{ncrpdf2}. Here too, 
for low energies the cosmic ray number density distribution has a long tail.

\subsection{Spatial intermittency of cosmic rays}
Cosmic rays fill all the volume available.
However, the distribution for low energy cosmic rays (especially for $\Lar/l_0 = 0.0016$ or 
$\Lar/l_b \simeq 0.06$) is not homogeneous.
Random magnetic fields 
produce cosmic ray distributions where a significant fraction of the volume is occupied
by strong particle concentrations. 
The PDFs of $\ncr$ shown in Fig.~\ref{ncrpdf1} have a Gaussian core and a heavy tail,
a manifestation of the spatial intermittency.
The core contains most of the particles and has $\ncr$ close to 
its mean value. The tail represents rare but intense small-scale
spatial structures. 
Figure~\ref{ncrconcone} shows the number density of cosmic rays, obtained as described 
in Section~\ref{TPSCR}, in the intermittent magnetic field
shown in the left-hand panel of \Fig{bvec} and \Fig{b2}. The distribution is inhomogeneous and evidently
sensitive to the magnetic field structure. 
The distribution of cosmic rays is affected by both the mean magnetic field and 
pitch angle scattering.
As shown in \Fig{ncrpdf2},
the mean magnetic field enhances the intermittency in 
the cosmic ray distribution, whereas
pitch angle scattering reduces it.
Fig.~\ref{ncrconcsec}
illustrates how the shape and number of structures in the cosmic ray distribution are affected.
With a mean field, the structures
are more numerous and many extend along the mean field direction, aligned with the
$x$-axis in the example shown.
This effect becomes significant when
$B_0/\brms \ge 1$. 
On the other hand, pitch angle scattering 
enhances cosmic ray diffusion and, consequently, reduces their intermittency, 
as shown in the right-hand panel of Fig.~\ref{ncrconcsec}.
The degree of intermittency can be measured in terms of
the parameter $f =  \langle \ncr \rangle^2/\langle \ncr^2 \rangle$
(see Table~\ref{table}).
For high energy particles, $\Lar/l_0 = 0.1592$ or 
$\Lar/l_b \simeq 6.52$,
$f$ is close to unity, indicating a homogeneous particle distribution.
This feature of the cosmic
ray distribution is further detailed in \Fig{ff} where we show the dependence 
of the fractional volume
of a region where $\ncr/\langle \ncr \rangle \ge \nu$
on $\nu$ for the configurations of Table~\ref{table}. 
Particles of sufficiently high energy, 
$\Lar/l_b \gg 1$, are not sensitive to the
fine structure of the magnetic field and the 
probability distribution of their number density is Gaussian.
The dependence of the fractional volume, shown in \Fig{ff}, 
on the presence of the mean magnetic field and its change due to pitch angle scattering 
confirms that the spatially intermittency increases on including mean field and decreases on including pitch angle scattering.
Figure~\ref{b2ncrpdf} shows the PDF of
magnetic field energy density and cosmic ray distribution for the 
tail region (values higher than the mean for each of the distributions). 
Both distributions are power laws and the exponent for the magnetic field roughly matches that  
for the cosmic ray number density. The cosmic ray distribution for low
energy particles is intermittent with heavy power law tails.

\subsection{Statistical relation between magnetic field and cosmic rays} \label{corr}
The simplest measure of a relation between
cosmic rays and magnetic field is their cross-correlation coefficient,
\begin{equation}\label{crosscorr}
	  C(\ncr,B^2) = 
	  \frac{\overline{\ncr B^2} 
	  	- \overline{n}_\mathrm{cr} \overline{B^2} }{\sigma_{\ncr} \sigma_{B^2}}\,,
\end{equation}
where the overbar denotes an average over the whole domain, and $\sigma$ is the standard 
deviation of the quantity specified in the subscript. The value of $C$ ranges from 
$C=1$ for perfect correlation to $C=-1$ for perfect anti-correlation.
In all cases considered, we find that the two distributions are uncorrelated,
$C\approx 0$.
This is true even for the lowest-energy cosmic rays considered ($\Lar/l_0=0.0016$),
despite the fact that they are closely confined to magnetic lines.
The correlation does not emerge even when the cosmic ray density and magnetic 
field are smoothed to a coarser spatial grid. To confirm that this behaviour is not 
an artefact of the initial conditions used for the cosmic ray particles,
we have also performed a simulation with their initial 
positions in the regions of the strongest magnetic field
shown in \Fig{b2}. The value of $C$ in this 
case is very close to unity initially but vanishes quickly, within 
the time $\tdiff$.

Figure~\ref{corr2d} shows the scatter plot of the cosmic ray number
density and magnetic field energy density, whose form confirms that 
the two variables are uncorrelated. We also confirm that cosmic 
rays and magnetic field distributions are statistically independent.
As demonstrated in Appendix~\ref{indep} (where more details can be found), the joint probability 
distribution function of cosmic rays and magnetic field distributions, $p(\ncr, b^2)$, 
can be factorized as follows: 
\begin{equation}
	p(\ncr,b)\approx(1.6+9.9
		\e^{-10.5 b^2/\brms^2}) \e^{-(\ncr/\langle\ncr\rangle-1)^2/0.18}\,,
\end{equation}
for the intermittent magnetic field and
\begin{equation}
	p(\ncr,b)\approx 0.5 (b^2/\brms^2)^{-1}\e^{-0.3 \ln(b^2/\brms^2)^2} \e^{-(\ncr/\langle\ncr\rangle-0.9)^2/0.32}\,,
\end{equation}
for the randomized (Gaussian) magnetic field. In both cases, 
the joint PDF is separable which illustrates that the cosmic rays and magnetic field distributions are independent
in the diffusive regime of the cosmic rays.
\begin{figure*} \centering
	\includegraphics[width=\columnwidth]{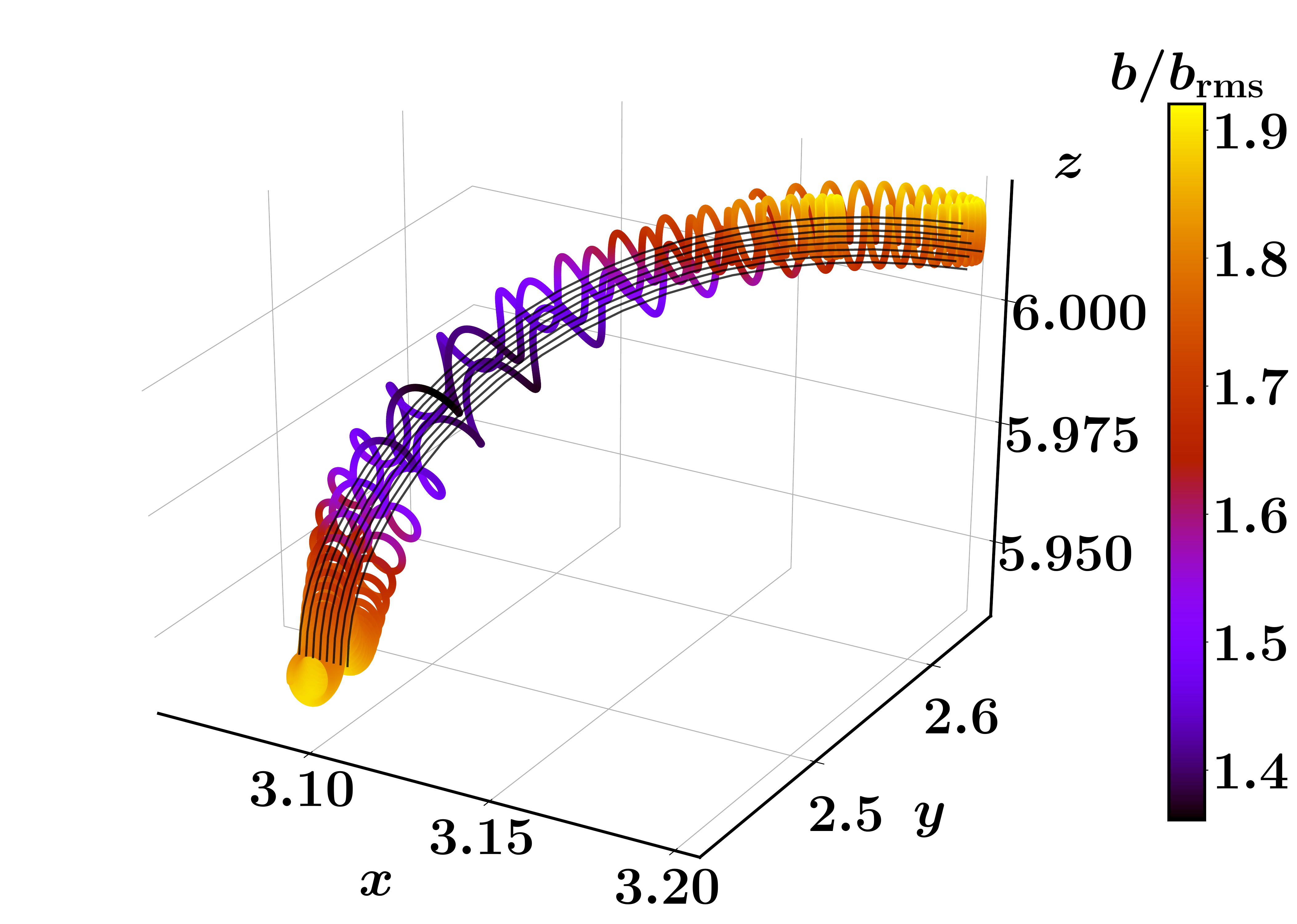}\qquad
	\includegraphics[width=0.9\columnwidth]{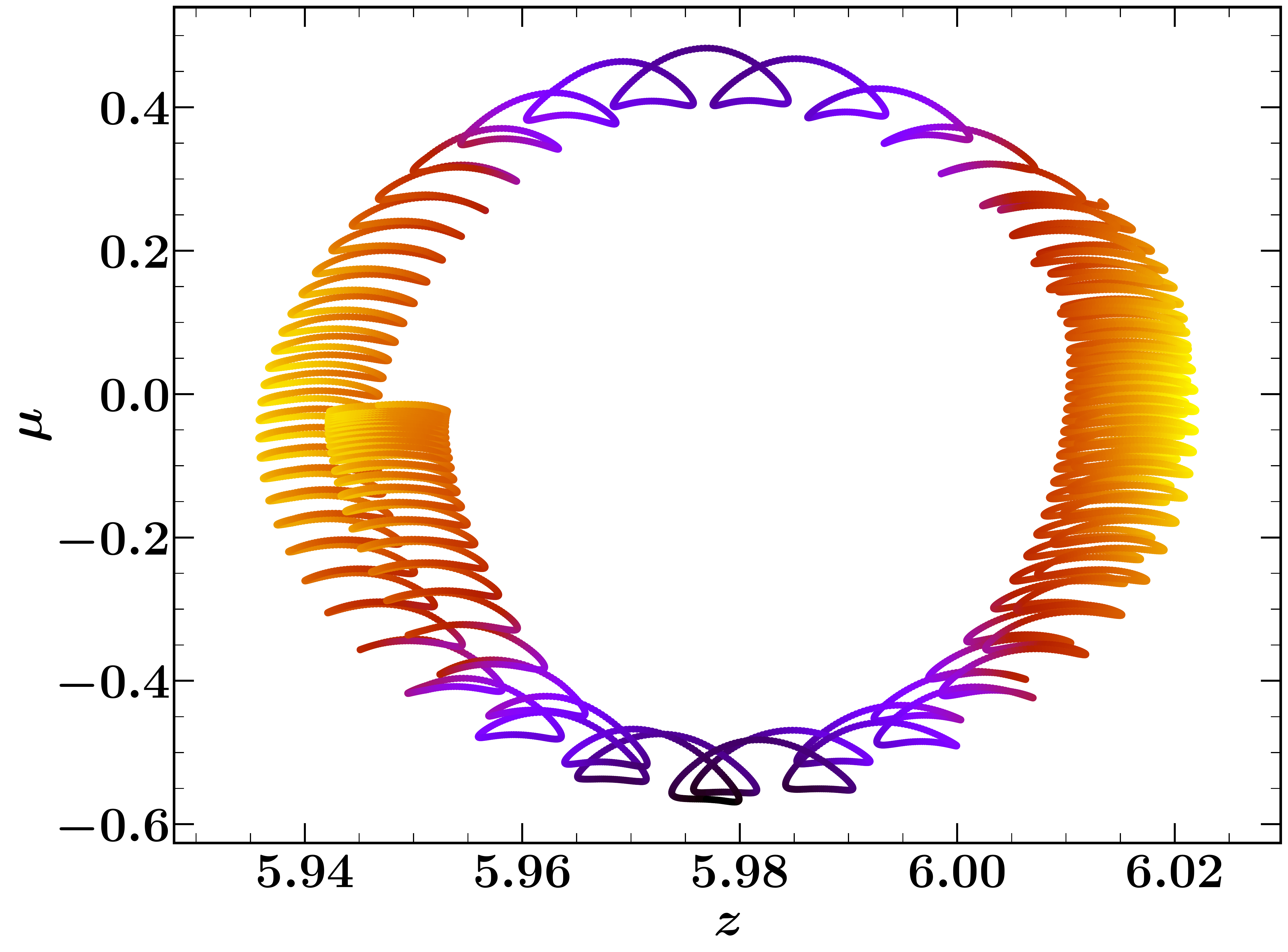}
	\caption{The left-hand panel shows a particle trajectory with magnetic field
		strength along the trajectory shown with colour. The dark grey lines show magnetic 
		field lines near the trajectory. The particle moves forward and backward
		between two magnetic mirrors. The right-hand panel shows the 
		angle between the particle velocity vector and the $z$-axis
		with magnetic field strength colour coded. The particle turns around 
		($\mu$ changes sign) at the magnetic mirrors, regions where magnetic field is 
		stronger.}
	\label{mir}
\end{figure*}

\subsection{Random magnetic traps}
Since $\ncr$ is not correlated with the magnetic field strength,
the regions of high cosmic ray density must instead be caused
by some geometrical property of the magnetic field lines.
In fact, we find that these regions occur where cosmic rays become trapped
between two magnetic mirrors,
i.e., positions where magnetic lines converge.
As shown in \Fig{mir},
if a magnetic flux tube is pinched at both ends,
then particles are repeatedly reflected
between the two ends,
creating a `magnetic trap'.
Because the field lines must be reasonably smooth in order to form a trap,
the regions of high cosmic ray density are typically smaller than the magnetic correlation length, $l_b$.
The cosine of the particle's pitch angle, $\mu$, defined in \Eq{mu}, is shown in the
right-hand panel of Fig.~\ref{mir} as a function of position and magnetic field 
strength. This quantity reverses sign along the trajectory whenever the particle 
is reflected at an end of the magnetic trap. This happens where the magnetic field 
is relatively strong. Magnetic trapping is associated with the conservation of $v_\perp^2/B$,
an adiabatic invariant \citep{Jackson}, where $v_\perp=v\sin\theta$ is
the particle speed perpendicular to the local magnetic field. We
have verified that $v_\perp^2/B=\text{const}$ with relative accuracy
of order $10^{-5}$ along the trajectories of the trapped particles.

We estimate the enhancement of cosmic ray number density due to magnetic traps as follows.
Consider a magnetic trap of a length $l$ in a magnetic flux tube of a radius $d$. 
For an ensemble of particles within the trap, the expected trapping time $\tau$ is $d^2/\kappa_\perp$, where $\kappa_\perp$ is the local transverse diffusivity of cosmic rays.
Defining $N$ to be the number of times that a particle travels along the trap before leaving it, we 
expect $N\sim \tau v/l$.
The resulting 
number density of particles within the trap is given by $n_0=N\overline{n}_\mathrm{cr}$
with  $\overline{n}_\mathrm{cr}$ the mean number density of cosmic rays.
The number density of the particles within the trap follows as
\begin{equation}\label{n0}
n_0\simeq\frac{v d^2}{l\kappa_\perp}\overline{n}_\mathrm{cr}.
\end{equation}
According to this order of magnitude estimate, we therefore expect that
the number of trapped particles will depend inversely upon the local cosmic ray diffusivity 
as well as upon the size and number of traps.

A mean magnetic field introduces a specific direction which particles follow
and so increases the probability for a magnetic trap to occur, as shown in
the left-hand panel of \Fig{ncrconcsec}. 
On the other hand, the pitch angle scattering due to self-generated waves
decreases the level of intermittency in the cosmic ray number density, as
shown in the right-hand panel of \Fig{ncrconcsec}, because
it facilitates the transverse diffusion.  When the distribution on the right-hand panel of Fig.~\ref{ncrconcsec}
is smoothed over a coarser scale, it does not correlate with the distribution on the 
left-hand panel of Fig.~\ref{ncrconcsec}. This further illustrates that the pitch angle scattering
not only reduces the length of trapping region but also changes the propagation of particles
significantly.

It is important to note that the spatial intermittency of cosmic rays 
does not require the intermittent structure of the random magnetic field.
Even in the randomized magnetic field, which has an almost perfect Gaussian statistics 
and is free of intermittency, magnetic traps occur and lead to a spatial intermittency
of cosmic rays.
\section{Conclusions and discussion}\label{conclusions}
Using test particle simulations of cosmic ray propagation, we have demonstrated that the spatial 
distribution of cosmic rays is not correlated with magnetic field strength on 
spatial scales that are less than (or comparable to) the 
outer scale, $l_0$, of a random flow that produces magnetic field. 
This implies that the local equipartition between cosmic ray and magnetic field energy 
densities does not occur on these scales. In fact, we find that cosmic ray number density
and magnetic field are approximately statistically independent.

At high energies, $\Lar > l_b$ with $l_b$ the correlation length of the magnetic field, the cosmic ray distribution is uniform. 
At low energies, $\Lar < l_b$, the spatial distribution of cosmic rays is 
intermittent in both Gaussian and non-Gaussian (spatially intermittent) magnetic fields.
This occurs because of the presence of magnetic traps in random magnetic fields
where the local magnetic field has a specific structure but is not necessarily
strong. As a result, the cosmic ray number density is not directly related to the 
local magnetic field strength.  

The trapping, and the ensuing intermittency in the cosmic ray distribution, are 
enhanced by a mean magnetic field and reduced by the pitch angle scattering of
cosmic ray particles or any other additional diffusion. The number density
of cosmic ray particles within the traps depends on the size of the trap
and the transverse diffusivity of cosmic rays with respect to the local magnetic field. 
Statistical properties
of the traps (such as their rate of occurrence, size, etc.) are controlled by
subtle properties of the random magnetic field; it appears that finding
them is a non-trivial probabilistic problem.

For cosmic rays to become trapped, their Larmor radius must be
smaller than the correlation length of the magnetic field, i.e., $\Lar \leq l_b$.
In spiral galaxies, where $l_b \lesssim 100 \pc$, the proton energy
in a $5 \mkG$ field with $\Lar = 100 \pc$ is $10^{8} \GeV$. Thus, the 
trapping of cosmic rays described above should be effective and efficient for 
galactic cosmic rays. 
In galaxy clusters, with $l_b\lesssim10\kpc$ and 
$b\simeq1\mkG$, protons of energy up to $10^9\GeV$ can be trapped and thus exhibit
spatial intermittency.
This particle trapping 
can effectively confine higher energy particles for which other confinement
mechanisms (scattering by self-generated waves, magnetic field structure, field line random walk)
are not very active \citep{Chandran2000}.

The lack of correlation between cosmic rays and magnetic field in galaxies at small 
scales (below a few kiloparsecs) is suggested by analyses of the radio--far-infrared 
correlation \citep{HBX98,BBT13,BR13}. Here we have presented a physical reason 
for that. Energy equipartition (or pressure balance) 
between cosmic rays and magnetic field may yet hold at
scales much larger than the driving scale of turbulence, $l_0\simeq100 \pc$
in spiral galaxies and $10 \kpc$ in galaxy clusters.

The assumption of energy equipartition between cosmic rays and magnetic fields,
and related assumptions of their pressure balance or minimum energy,
are routinely used in interpretations of synchrotron observations
\citep[][and references therein]{Beck2005,AUAPV12,Beck2016}. Apart from
assuming that the cosmic ray energy density, dominated by protons, is
related to the local magnetic energy density, these interpretations also rely
on a fixed ratio of the number densities of cosmic ray electrons and protons and
presume that the electrons and protons are identically distributed in space. We have 
shown that \textit{local}, small-scale energy equipartition between cosmic rays and 
random magnetic field does not occur. The spatial distribution of cosmic
ray energy is sensitive to the geometry of magnetic field rather than its strength.
However, the distribution of cosmic ray electrons 
responsible for synchrotron emission must be modelled separately with allowance for 
their synchrotron and inverse-Compton energy losses. We expect that the 
distribution of cosmic ray electrons would differ significantly from that of protons, 
plausibly exhibiting stronger intermittency since the electrons would 
spend more time in magnetic traps as they 
lose their energy. This would mean 
that equipartition-type estimates are even less reliable when applied to the local 
energy densities. 
It cannot be excluded, however, that the spatial distributions of cosmic
ray and magnetic field energy densities are related at larger scales of order
kiloparsec, comparable to the diffusion length of cosmic ray particles over
their confinement time. We shall discuss elsewhere this version of the equipartition arguments.

The intermittency in the distribution of cosmic ray particles discussed above
is a robust feature of their propagation in a random magnetic field that emerges
in both Gaussian and non-Gaussian magnetic fields. This extreme small-scale
inhomogeneity may affect interactions of cosmic rays with interstellar gas
\citep[][and references therein]{GBS15}, including spallation reactions and 
ionization, e.g., producing strong local variations in the ionization rate. The 
$\gamma$-ray emissivity can also be affected by the trapping of cosmic rays 	
particles in random magnetic traps. Quantifying the effect of cosmic ray intermittency on these and other observables (such as spectra)
will require energy losses to be included in the model, which we aim to do in the future.The effects of intermittency in the cosmic ray 
distribution may be numerous and require a careful, systematic study.

Test particle simulations are constrained by the need to resolve the Larmor 
radius and period of the particle gyration.
For this reason, it is not practical
to model those particles of energies of order GeV that contribute most to the 
energy density of cosmic rays if the magnetic field structure is to be resolved
in full. For example, the Larmor radius of a $5 \GeV$ proton in a $5 \mkG$ ISM magnetic field 
is $10^8$ times smaller than $100 \pc$, the typical scale of the magnetic field. Our conclusions are based on an extrapolation of the results obtained for 
particles of effectively much larger energies. Then modelling of cosmic ray propagation at lower energies
requires a fluid model based on a variant of the diffusion-advection equation but
with the diffusion tensor that allows for the non-trivial structure of the magnetic 
field. Such modelling is beyond the scope of this paper but remains our high priority.

\section{Acknowledgments}
We are grateful to Ellen Zweibel, Prateek Sharma and Pasquale Blasi for useful discussions. We thank the anonymous referee for her/his helpful suggestions and comments.
This work was supported by the STFC (ST/N000900/1, Project 2), the Leverhulme Trust (RPG-2014-427) and the Thailand Research Fund (RTA5980003).

\bibliographystyle{mnras}
\bibliography{eq}

\begin{thebibliography}{}
\makeatletter
\relax
\def\mn@urlcharsother{\let\do\@makeother \do\$\do\&\do\#\do\^\do\_\do\%\do\~}
\def\mn@doi{\begingroup\mn@urlcharsother \@ifnextchar [ {\mn@doi@}
  {\mn@doi@[]}}
\def\mn@doi@[#1]#2{\def\@tempa{#1}\ifx\@tempa\@empty \href
  {http://dx.doi.org/#2} {doi:#2}\else \href {http://dx.doi.org/#2} {#1}\fi
  \endgroup}
\def\mn@eprint#1#2{\mn@eprint@#1:#2::\@nil}
\def\mn@eprint@arXiv#1{\href {http://arxiv.org/abs/#1} {{\tt arXiv:#1}}}
\def\mn@eprint@dblp#1{\href {http://dblp.uni-trier.de/rec/bibtex/#1.xml}
  {dblp:#1}}
\def\mn@eprint@#1:#2:#3:#4\@nil{\def\@tempa {#1}\def\@tempb {#2}\def\@tempc
  {#3}\ifx \@tempc \@empty \let \@tempc \@tempb \let \@tempb \@tempa \fi \ifx
  \@tempb \@empty \def\@tempb {arXiv}\fi \@ifundefined
  {mn@eprint@\@tempb}{\@tempb:\@tempc}{\expandafter \expandafter \csname
  mn@eprint@\@tempb\endcsname \expandafter{\@tempc}}}

\bibitem[\protect\citeauthoryear{{Arbutina}, {Uro{\v s}evi{\'c}},
  {Andjeli{\'c}}, {Pavlovi{\'c}}  \& {Vukoti{\'c}}}{{Arbutina}
  et~al.}{2012}]{AUAPV12}
{Arbutina} B.,  {Uro{\v s}evi{\'c}} D.,  {Andjeli{\'c}} M.~M.,  {Pavlovi{\'c}}
  M.~Z.,   {Vukoti{\'c}} B.,  2012, \mn@doi [\apj]
  {10.1088/0004-637X/746/1/79}, \href
  {http://adsabs.harvard.edu/abs/2012ApJ...746...79A} {746, 79}

\bibitem[\protect\citeauthoryear{{Basu} \& {Roy}}{{Basu} \& {Roy}}{2013}]{BR13}
{Basu} A.,  {Roy} S.,  2013, \mn@doi [\mnras] {10.1093/mnras/stt845}, \href
  {http://adsabs.harvard.edu/abs/2013MNRAS.433.1675B} {433, 1675}

\bibitem[\protect\citeauthoryear{{Beck}}{{Beck}}{2016}]{Beck2016}
{Beck} R.,  2016, \mn@doi [\aapr] {10.1007/s00159-015-0084-4}, \href
  {http://adsabs.harvard.edu/abs/2016A%26ARv..24....4B} {24, 4}

\bibitem[\protect\citeauthoryear{{Beck} \& {Krause}}{{Beck} \&
  {Krause}}{2005}]{Beck2005}
{Beck} R.,  {Krause} M.,  2005, \mn@doi [\an] {10.1002/asna.200510366}, \href
  {http://adsabs.harvard.edu/abs/2005AN....326..414B} {326, 414}

\bibitem[\protect\citeauthoryear{{Beck}, {Brandenburg}, {Moss}, {Shukurov}  \&
  {Sokoloff}}{{Beck} et~al.}{1996}]{Beck1996}
{Beck} R.,  {Brandenburg} A.,  {Moss} D.,  {Shukurov} A.,   {Sokoloff} D.,
  1996, \mn@doi [\araa] {10.1146/annurev.astro.34.1.155}, \href
  {http://adsabs.harvard.edu/abs/1996ARA%26A..34..155B} {34, 155}

\bibitem[\protect\citeauthoryear{{Beck}, {Shukurov}, {Sokoloff}  \&
  {Wielebinski}}{{Beck} et~al.}{2003}]{Beck2003}
{Beck} R.,  {Shukurov} A.,  {Sokoloff} D.,   {Wielebinski} R.,  2003, \mn@doi
  [\aap] {10.1051/0004-6361:20031101}, \href
  {http://adsabs.harvard.edu/abs/2003A%26A...411...99B} {411, 99}

\bibitem[\protect\citeauthoryear{{Berezinskii}, {Bulanov}, {Dogiel}  \&
  {Ptuskin}}{{Berezinskii} et~al.}{1990}]{BBDP90}
{Berezinskii} V.~S.,  {Bulanov} S.~V.,  {Dogiel} V.~A.,   {Ptuskin} V.~S.,
  1990, {Astrophysics of Cosmic Rays}.
North-Holland, Amsterdam

\bibitem[\protect\citeauthoryear{{Berkhuijsen}, {Beck}  \&
  {Tabatabaei}}{{Berkhuijsen} et~al.}{2013}]{BBT13}
{Berkhuijsen} E.~M.,  {Beck} R.,   {Tabatabaei} F.~S.,  2013, \mn@doi [\mnras]
  {10.1093/mnras/stt1400}, \href
  {http://adsabs.harvard.edu/abs/2013MNRAS.435.1598B} {435, 1598}

\bibitem[\protect\citeauthoryear{{Biskamp}}{{Biskamp}}{2003}]{Biskamp2003}
{Biskamp} D.,  2003, {Magnetohydrodynamic Turbulence}.
Cambridge University Press, Cambridge, UK

\bibitem[\protect\citeauthoryear{{Brandenburg} \& {Subramanian}}{{Brandenburg}
  \& {Subramanian}}{2005}]{BS2005}
{Brandenburg} A.,  {Subramanian} K.,  2005, \mn@doi [\physrep]
  {10.1016/j.physrep.2005.06.005}, \href
  {http://adsabs.harvard.edu/abs/2005PhR...417....1B} {417, 1}

\bibitem[\protect\citeauthoryear{{Burbidge}}{{Burbidge}}{1956}]{B56}
{Burbidge} G.~R.,  1956, \mn@doi [\apj] {10.1086/146237}, \href
  {http://adsabs.harvard.edu/abs/1956ApJ...124..416B} {124, 416}

\bibitem[\protect\citeauthoryear{{Bykov}}{{Bykov}}{1988}]{B88}
{Bykov} A.~M.,  1988, \sal, \href
  {http://adsabs.harvard.edu/abs/1988SvAL...14...60B} {14, 60}

\bibitem[\protect\citeauthoryear{{Bykov} \& {Toptygin}}{{Bykov} \&
  {Toptygin}}{1985}]{BT85}
{Bykov} A.~M.,  {Toptygin} I.~N.,  1985, \sal, \href
  {http://adsabs.harvard.edu/abs/1985SvAL...11...75B} {11, 75}

\bibitem[\protect\citeauthoryear{{Bykov} \& {Toptygin}}{{Bykov} \&
  {Toptygin}}{1987}]{ByT87}
{Bykov} A.~M.,  {Toptygin} I.~N.,  1987, \mn@doi [\apss] {10.1007/BF00637855},
  \href {http://adsabs.harvard.edu/abs/1987Ap%26SS.138..341B} {138, 341}

\bibitem[\protect\citeauthoryear{{Candia} \& {Roulet}}{{Candia} \&
  {Roulet}}{2004}]{CandiaRoulet2004}
{Candia} J.,  {Roulet} E.,  2004, \mn@doi [\jcap]
  {10.1088/1475-7516/2004/10/007}, \href
  {http://adsabs.harvard.edu/abs/2004JCAP...10..007C} {10, 7}

\bibitem[\protect\citeauthoryear{{Casse}, {Lemoine}  \& {Pelletier}}{{Casse}
  et~al.}{2002}]{Casse_et_al2002}
{Casse} F.,  {Lemoine} M.,   {Pelletier} G.,  2002, \mn@doi [\prd]
  {10.1103/PhysRevD.65.023002}, \href
  {http://adsabs.harvard.edu/abs/2002PhRvD..65b3002C} {65, 023002}

\bibitem[\protect\citeauthoryear{{Cesarsky} \& {Kulsrud}}{{Cesarsky} \&
  {Kulsrud}}{1981}]{CK81}
{Cesarsky} C.~J.,  {Kulsrud} R.~M.,  1981, in {Setti} G.,  {Spada} G.,
  {Wolfendale} A.~W.,  eds,  IAU Symposium Vol. 94, Origin of Cosmic Rays.
  p.~251

\bibitem[\protect\citeauthoryear{{Chamandy}, {Shukurov}, {Subramanian}  \&
  {Stoker}}{{Chamandy} et~al.}{2014}]{Chamandy2014}
{Chamandy} L.,  {Shukurov} A.,  {Subramanian} K.,   {Stoker} K.,  2014, \mn@doi
  [\mnras] {10.1093/mnras/stu1274}, \href
  {http://adsabs.harvard.edu/abs/2014MNRAS.443.1867C} {443, 1867}

\bibitem[\protect\citeauthoryear{{Chandran}}{{Chandran}}{2000}]{Chandran2000}
{Chandran} B.~D.~G.,  2000, \mn@doi [\apj] {10.1086/308232}, \href
  {http://adsabs.harvard.edu/abs/2000ApJ...529..513C} {529, 513}

\bibitem[\protect\citeauthoryear{{DeMarco}, {Blasi}  \& {Stanev}}{{DeMarco}
  et~al.}{2007}]{DBS2007}
{DeMarco} D.,  {Blasi} P.,   {Stanev} T.,  2007, \mn@doi [\jcap]
  {10.1088/1475-7516/2007/06/027}, \href
  {http://adsabs.harvard.edu/abs/2007JCAP...06..027D} {6, 027}

\bibitem[\protect\citeauthoryear{{Desiati} \& {Zweibel}}{{Desiati} \&
  {Zweibel}}{2014}]{DZ2014}
{Desiati} P.,  {Zweibel} E.~G.,  2014, \mn@doi [\apj]
  {10.1088/0004-637X/791/1/51}, \href
  {http://adsabs.harvard.edu/abs/2014ApJ...791...51D} {791, 51}

\bibitem[\protect\citeauthoryear{{Evirgen}, {Gent}, {Shukurov}, {Fletcher}  \&
  {Bushby}}{{Evirgen} et~al.}{2017}]{EGSFB17}
{Evirgen} C.~C.,  {Gent} F.~A.,  {Shukurov} A.,  {Fletcher} A.,   {Bushby} P.,
  2017, \mn@doi [\mnras] {10.1093/mnrasl/slw196}, \href
  {http://adsabs.harvard.edu/abs/2017MNRAS.464L.105E} {464, L105}

\bibitem[\protect\citeauthoryear{{Farmer} \& {Goldreich}}{{Farmer} \&
  {Goldreich}}{2004}]{FG2004}
{Farmer} A.~J.,  {Goldreich} P.,  2004, \mn@doi [\apj] {10.1086/382040}, \href
  {http://adsabs.harvard.edu/abs/2004ApJ...604..671F} {604, 671}

\bibitem[\protect\citeauthoryear{{Federrath}, {Roman-Duval}, {Klessen},
  {Schmidt}  \& {Mac Low}}{{Federrath} et~al.}{2010}]{Federrath2010}
{Federrath} C.,  {Roman-Duval} J.,  {Klessen} R.~S.,  {Schmidt} W.,   {Mac Low}
  M.-M.,  2010, \mn@doi [\aap] {10.1051/0004-6361/200912437}, \href
  {http://adsabs.harvard.edu/abs/2010A%26A...512A..81F} {512, A81}

\bibitem[\protect\citeauthoryear{{Felice} \& {Kulsrud}}{{Felice} \&
  {Kulsrud}}{2001}]{FK2001}
{Felice} G.~M.,  {Kulsrud} R.~M.,  2001, \mn@doi [\apj] {10.1086/320651}, \href
  {http://adsabs.harvard.edu/abs/2001ApJ...553..198F} {553, 198}

\bibitem[\protect\citeauthoryear{{Fung}, {Hunt}, {Malik}  \& {Perkins}}{{Fung}
  et~al.}{1992}]{FungEA1992}
{Fung} J.~C.~H.,  {Hunt} J.~C.~R.,  {Malik} N.~A.,   {Perkins} R.~J.,  1992,
  \mn@doi [\jfm] {10.1017/S0022112092001423}, 236, 281

\bibitem[\protect\citeauthoryear{{Gaensler} et~al.,}{{Gaensler}
  et~al.}{2011}]{Gaenslar2011}
{Gaensler} B.~M.,  et~al., 2011, \mn@doi [\nat] {10.1038/nature10446}, \href
  {http://adsabs.harvard.edu/abs/2011Natur.478..214G} {478, 214}

\bibitem[\protect\citeauthoryear{{Giacalone} \& {Jokipii}}{{Giacalone} \&
  {Jokipii}}{1999}]{GJ1999}
{Giacalone} J.,  {Jokipii} J.~R.,  1999, \mn@doi [\apj] {10.1086/307452}, \href
  {http://adsabs.harvard.edu/abs/1999ApJ...520..204G} {520, 204}

\bibitem[\protect\citeauthoryear{{Globus}, {Allard}  \& {Parizot}}{{Globus}
  et~al.}{2008}]{GAP08}
{Globus} N.,  {Allard} D.,   {Parizot} E.,  2008, \mn@doi [\aap]
  {10.1051/0004-6361:20078653}, \href
  {http://adsabs.harvard.edu/abs/2008A%26A...479...97G} {479, 97}

\bibitem[\protect\citeauthoryear{{Grenier}, {Black}  \& {Strong}}{{Grenier}
  et~al.}{2015}]{GBS15}
{Grenier} I.~A.,  {Black} J.~H.,   {Strong} A.~W.,  2015, \mn@doi [\araa]
  {10.1146/annurev-astro-082214-122457}, \href
  {http://adsabs.harvard.edu/abs/2015ARA%26A..53..199G} {53, 199}

\bibitem[\protect\citeauthoryear{{Harari}, {Mollerach}  \& {Roulet}}{{Harari}
  et~al.}{2014}]{HMR14}
{Harari} D.,  {Mollerach} S.,   {Roulet} E.,  2014, \mn@doi [\prd]
  {10.1103/PhysRevD.89.123001}, \href
  {http://adsabs.harvard.edu/abs/2014PhRvD..89l3001H} {89, 123001}

\bibitem[\protect\citeauthoryear{{Haverkorn} \& {Spangler}}{{Haverkorn} \&
  {Spangler}}{2013}]{HS13}
{Haverkorn} M.,  {Spangler} S.~R.,  2013, \mn@doi [\ssr]
  {10.1007/s11214-013-0014-6}, \href
  {http://adsabs.harvard.edu/abs/2013SSRv..178..483H} {178, 483}

\bibitem[\protect\citeauthoryear{{Heiles} \& {Troland}}{{Heiles} \&
  {Troland}}{2005}]{HT05}
{Heiles} C.,  {Troland} T.~H.,  2005, \mn@doi [\apj] {10.1086/428896}, \href
  {http://adsabs.harvard.edu/abs/2005ApJ...624..773H} {624, 773}

\bibitem[\protect\citeauthoryear{{Hoernes}, {Berkhuijsen}  \& {Xu}}{{Hoernes}
  et~al.}{1998}]{HBX98}
{Hoernes} P.,  {Berkhuijsen} E.~M.,   {Xu} C.,  1998, \aap, \href
  {http://adsabs.harvard.edu/abs/1998A%26A...334...57H} {334, 57}

\bibitem[\protect\citeauthoryear{{Jackson}}{{Jackson}}{1998}]{Jackson}
{Jackson} J.~D.,  1998, {Classical Electrodynamics, 3rd Edition}.
{Wiley}, {New York}

\bibitem[\protect\citeauthoryear{{Kalberla} \& {Kerp}}{{Kalberla} \&
  {Kerp}}{2016}]{KK2016}
{Kalberla} P.~M.~W.,  {Kerp} J.,  2016, \mn@doi [\aap]
  {10.1051/0004-6361/201629113}, \href
  {http://adsabs.harvard.edu/abs/2016A%26A...595A..37K} {595, A37}

\bibitem[\protect\citeauthoryear{Kazantsev}{Kazantsev}{1967}]{Kaz68}
Kazantsev A.~P.,  1967, \zhetp, 53, 1807

\bibitem[\protect\citeauthoryear{{Klein} \& {Fletcher}}{{Klein} \&
  {Fletcher}}{2015}]{KleinFletcher2015}
{Klein} U.,  {Fletcher} A.,  2015, {Galactic and Intergalactic Magnetic
  Fields}.
{Springer Praxis Books, Springer International Publishing, Heidelberg, Germany}

\bibitem[\protect\citeauthoryear{{Kulsrud}}{{Kulsrud}}{2005}]{Kulsrud2005}
{Kulsrud} R.~M.,  2005, {Plasma Physics for Astrophysics}.
{Princeton University Press, Princeton}

\bibitem[\protect\citeauthoryear{{Kulsrud} \& {Pearce}}{{Kulsrud} \&
  {Pearce}}{1969}]{KP1969}
{Kulsrud} R.,  {Pearce} W.~P.,  1969, \mn@doi [\apj] {10.1086/149981}, \href
  {http://adsabs.harvard.edu/abs/1969ApJ...156..445K} {156, 445}

\bibitem[\protect\citeauthoryear{{Michalek} \& {Ostrowski}}{{Michalek} \&
  {Ostrowski}}{1997}]{MichalekOstrowski97}
{Michalek} G.,  {Ostrowski} M.,  1997, \aap, \href
  {http://adsabs.harvard.edu/abs/1997A%26A...326..793M} {326, 793}

\bibitem[\protect\citeauthoryear{{Moffatt}}{{Moffatt}}{1978}]{Moffatt1980}
{Moffatt} H.~K.,  1978, {Magnetic field generation in electrically conducting
  fluids.}.
Cambridge University Press, Cambridge

\bibitem[\protect\citeauthoryear{{Monin} \& {Yaglom}}{{Monin} \&
  {Yaglom}}{1971}]{Monin_Yaglom1971}
{Monin} A.~S.,  {Yaglom} A.~M.,  1971, {Statistical Fluid Mechanics}.
MIT Press, Cambridge, Massachusetts, USA

\bibitem[\protect\citeauthoryear{{Moss}, {Snodin}, {Englmaier}, {Shukurov},
  {Beck}  \& {Sokoloff}}{{Moss} et~al.}{2007}]{Moss2007}
{Moss} D.,  {Snodin} A.~P.,  {Englmaier} P.,  {Shukurov} A.,  {Beck} R.,
  {Sokoloff} D.~D.,  2007, \mn@doi [\aap] {10.1051/0004-6361:20066222}, \href
  {http://adsabs.harvard.edu/abs/2007A%26A...465..157M} {465, 157}

\bibitem[\protect\citeauthoryear{{Parizot}}{{Parizot}}{2004}]{Parizot2004}
{Parizot} E.,  2004, \mn@doi [Nucl.\ Phys.\ B, Proc.\ Suppl.]
  {10.1016/j.nuclphysbps.2004.10.034}, \href
  {http://adsabs.harvard.edu/abs/2004NuPhS.136..169P} {136, 169}

\bibitem[\protect\citeauthoryear{{Plotnikov}, {Pelletier}  \&
  {Lemoine}}{{Plotnikov} et~al.}{2011}]{Plotnikov_et_al2011}
{Plotnikov} I.,  {Pelletier} G.,   {Lemoine} M.,  2011, \mn@doi [\aap]
  {10.1051/0004-6361/201117182}, \href
  {http://adsabs.harvard.edu/abs/2011A%26A...532A..68P} {532, A68}

\bibitem[\protect\citeauthoryear{{Ruzmaikin}, {Sokoloff}  \&
  {Shukurov}}{{Ruzmaikin} et~al.}{1989}]{RSS89}
{Ruzmaikin} A.,  {Sokoloff} D.,   {Shukurov} A.,  1989, \mn@doi [\mnras]
  {10.1093/mnras/241.1.1}, \href
  {http://adsabs.harvard.edu/abs/1989MNRAS.241....1R} {241, 1}

\bibitem[\protect\citeauthoryear{{Schekochihin}, {Cowley}, {Taylor}, {Maron}
  \& {McWilliams}}{{Schekochihin} et~al.}{2004}]{SCTMM04}
{Schekochihin} A.~A.,  {Cowley} S.~C.,  {Taylor} S.~F.,  {Maron} J.~L.,
  {McWilliams} J.~C.,  2004, \mn@doi [\apj] {10.1086/422547}, \href
  {http://adsabs.harvard.edu/abs/2004ApJ...612..276S} {612, 276}

\bibitem[\protect\citeauthoryear{{Schekochihin}, {Cowley}, {Dorland},
  {Hammett}, {Howes}, {Quataert}  \& {Tatsuno}}{{Schekochihin}
  et~al.}{2009}]{SCDHHQT09}
{Schekochihin} A.~A.,  {Cowley} S.~C.,  {Dorland} W.,  {Hammett} G.~W.,
  {Howes} G.~G.,  {Quataert} E.,   {Tatsuno} T.,  2009, \mn@doi [\apjs]
  {10.1088/0067-0049/182/1/310}, \href
  {http://adsabs.harvard.edu/abs/2009ApJS..182..310S} {182, 310}

\bibitem[\protect\citeauthoryear{{Schlickeiser}}{{Schlickeiser}}{2002}]{Sch02}
{Schlickeiser} R.,  2002, {Cosmic Ray Astrophysics}.
Springer, Berlin

\bibitem[\protect\citeauthoryear{{Shalchi}}{{Shalchi}}{2009}]{Sh09}
{Shalchi} A.,  2009, {Nonlinear Cosmic Ray Diffusion Theories}.
~ Vol. 362, Springer, Berlin

\bibitem[\protect\citeauthoryear{{Shukurov}, {Snodin}, {Seta}, {Bushby}  \&
  {Wood}}{{Shukurov} et~al.}{2017}]{SSSBW17}
{Shukurov} A.,  {Snodin} A.~P.,  {Seta} A.,  {Bushby} P.~J.,   {Wood} T.~S.,
  2017, \mn@doi [\apjl] {10.3847/2041-8213/aa6aa6}, \href
  {http://adsabs.harvard.edu/abs/2017ApJ...839L..16S} {839, L16}

\bibitem[\protect\citeauthoryear{{Snodin}, {Brandenburg}, {Mee}  \&
  {Shukurov}}{{Snodin} et~al.}{2006}]{SBMS06}
{Snodin} A.~P.,  {Brandenburg} A.,  {Mee} A.~J.,   {Shukurov} A.,  2006,
  \mn@doi [\mnras] {10.1111/j.1365-2966.2006.11034.x}, \href
  {http://adsabs.harvard.edu/abs/2006MNRAS.373..643S} {373, 643}

\bibitem[\protect\citeauthoryear{{Snodin}, {Ruffolo}, {Oughton}, {Servidio}  \&
  {Matthaeus}}{{Snodin} et~al.}{2013}]{Snodin2013}
{Snodin} A.~P.,  {Ruffolo} D.,  {Oughton} S.,  {Servidio} S.,   {Matthaeus}
  W.~H.,  2013, \mn@doi [\apj] {10.1088/0004-637X/779/1/56}, \href
  {http://adsabs.harvard.edu/abs/2013ApJ...779...56S} {779, 56}

\bibitem[\protect\citeauthoryear{{Snodin}, {Shukurov}, {Sarson}, {Bushby}  \&
  {Rodrigues}}{{Snodin} et~al.}{2016}]{Snodin_et_al2016}
{Snodin} A.~P.,  {Shukurov} A.,  {Sarson} G.~R.,  {Bushby} P.~J.,   {Rodrigues}
  L.~F.~S.,  2016, \mn@doi [\mnras] {10.1093/mnras/stw217}, \href
  {http://adsabs.harvard.edu/abs/2016MNRAS.457.3975S} {457, 3975}

\bibitem[\protect\citeauthoryear{{Stepanov}, {Shukurov}, {Fletcher}, {Beck},
  {La Porta}  \& {Tabatabaei}}{{Stepanov} et~al.}{2014}]{Stepanov2014}
{Stepanov} R.,  {Shukurov} A.,  {Fletcher} A.,  {Beck} R.,  {La Porta} L.,
  {Tabatabaei} F.,  2014, \mn@doi [\mnras] {10.1093/mnras/stt2044}, \href
  {http://adsabs.harvard.edu/abs/2014MNRAS.437.2201S} {437, 2201}

\bibitem[\protect\citeauthoryear{{Strong}, {Moskalenko}  \& {Ptuskin}}{{Strong}
  et~al.}{2007}]{Strong2007}
{Strong} A.~W.,  {Moskalenko} I.~V.,   {Ptuskin} V.~S.,  2007, \mn@doi [Annual
  Review of Nuclear and Particle Science]
  {10.1146/annurev.nucl.57.090506.123011}, \href
  {http://adsabs.harvard.edu/abs/2007ARNPS..57..285S} {57, 285}

\bibitem[\protect\citeauthoryear{{Subedi} et~al.,}{{Subedi}
  et~al.}{2017}]{Subedi+17}
{Subedi} P.,  et~al., 2017, \mn@doi [\apj] {10.3847/1538-4357/aa603a}, \href
  {http://adsabs.harvard.edu/abs/2017ApJ...837..140S} {837, 140}

\bibitem[\protect\citeauthoryear{{Subramanian}}{{Subramanian}}{1999}]{Sub99}
{Subramanian} K.,  1999, \mn@doi [\prl] {10.1103/PhysRevLett.83.2957}, \href
  {http://adsabs.harvard.edu/abs/1999PhRvL..83.2957S} {83, 2957}

\bibitem[\protect\citeauthoryear{{Subramanian}, {Shukurov}  \&
  {Haugen}}{{Subramanian} et~al.}{2006}]{SSH06}
{Subramanian} K.,  {Shukurov} A.,   {Haugen} N.~E.~L.,  2006, \mn@doi [\mnras]
  {10.1111/j.1365-2966.2006.09918.x}, \href
  {http://adsabs.harvard.edu/abs/2006MNRAS.366.1437S} {366, 1437}

\bibitem[\protect\citeauthoryear{{Wentzel}}{{Wentzel}}{1974}]{Wentzel1974}
{Wentzel} D.~G.,  1974, \mn@doi [\araa] {10.1146/annurev.aa.12.090174.000443},
  \href {http://adsabs.harvard.edu/abs/1974ARA%26A..12...71W} {12, 71}

\bibitem[\protect\citeauthoryear{{Wilkin}, {Barenghi}  \& {Shukurov}}{{Wilkin}
  et~al.}{2007}]{Wilkin2007}
{Wilkin} S.~L.,  {Barenghi} C.~F.,   {Shukurov} A.,  2007, \mn@doi [\prl]
  {10.1103/PhysRevLett.99.134501}, \href
  {http://adsabs.harvard.edu/abs/2007PhRvL..99m4501W} {99, 134501}

\bibitem[\protect\citeauthoryear{{Zaroubi} et~al.,}{{Zaroubi}
  et~al.}{2015}]{Z+15}
{Zaroubi} S.,  et~al., 2015, \mn@doi [\mnras] {10.1093/mnrasl/slv123}, \href
  {http://adsabs.harvard.edu/abs/2015MNRAS.454L..46Z} {454, L46}

\bibitem[\protect\citeauthoryear{{Zeldovich}, {Ruzmaikin}  \&
  {Sokoloff}}{{Zeldovich} et~al.}{1990}]{ZRS90}
{Zeldovich} {\relax Ya}.~B.,  {Ruzmaikin} A.~A.,   {Sokoloff} D.~D.,  1990,
  {The Almighty Chance}.
World Scientific, Singapore

\bibitem[\protect\citeauthoryear{{Zelenyi}, {Bykov}, {Uvarov}  \&
  {Artemyev}}{{Zelenyi} et~al.}{2015}]{ZBUA15}
{Zelenyi} L.~M.,  {Bykov} A.~M.,  {Uvarov} Y.~A.,   {Artemyev} A.~V.,  2015,
  \mn@doi [\jpp] {10.1017/S0022377815000409}, \href
  {http://adsabs.harvard.edu/abs/2015JPlPh..81d3901Z} {81, 395810401}

\makeatother
\end{thebibliography}

\appendix

\section{Fine structure of interstellar magnetic fields} \label{finescale}
Several mechanisms produce
the random magnetic fields in the interstellar medium. Tangling of the
large-scale magnetic field by turbulent flows produces a volume-filling random field whose statistical properties
are controlled by the turbulence. This magnetic field is a byproduct of the large-scale (mean-field) dynamo action.
If the turbulent velocity has Gaussian statistical properties, the resulting random magnetic field is also a
Gaussian random field. Random motions can also generate a random magnetic field directly through the small-scale (fluctuation) dynamo action. The resulting magnetic field
is spatially intermittent (magnetic field concentrated in filaments and ribbons) and has strongly non-Gaussian statistical properties 
\citep{ZRS90,SCTMM04,BS2005,Wilkin2007}. 
Another similarly structured
contribution is produced by compression in random shock fronts driven
by supernova explosions \citep{BT85, Federrath2010}. The result of such compression is a complex random magnetic field 
represented by both Gaussian and non-Gaussian parts that scatter cosmic rays differently \citep{ZBUA15,SSSBW17}. 
In this section we estimate the contribution of each of the above mechanisms to the small-scale interstellar magnetic fields. 
For clarity, we denote $\vec{b}_1$, $\vec{b}_2$ and $\vec{b}_3$ the random magnetic fields produced
by tangling of the mean field, by the fluctuation dynamo action and by shock compression, respectively.

The tangling of a large-scale magnetic field $\vec{B_0}$ by a random flow $\vec{u}$ can be 
described by
the induction equation, \Eq{indeqn}, with magnetic diffusion 
neglected over the time-scales of interest,
$\partial\vec{b}_1/\partial t \approx \nabla\times(\vec{u}\times\vec{B_0})$
\citep[e.g.,][]{Moffatt1980}. 
By order of magnitude, $b_1\simeq uB_0\tau/l$, where $\tau$ and $l$ are
the relevant time and length scales. Assuming that $\tau$ is equal to the eddy turnover time, $l/u$,
we obtain $b_1 \simeq B_0$.
This part of the random magnetic field is present wherever $B_0\neq0$, i.e., presumably at all positions.
The strength of the interstellar large-scale magnetic field is controlled by the global properties of ISM, such as
the velocity shear rate due to differential rotation, the Rossby number and the magnetic helicity flux 
\citep[see][for a compilation of useful results for nonlinear mean-field dynamos]{Chamandy2014}.
Observations suggests $B_0 \simeq 1\text{--}5 \mkG$, varying between
galaxies and between various locations within a given galaxy  \citep{Beck2016}. Thus $b_1\simeq 1\text{--}5 \mkG$.

Magnetic structures produced by the small-scale dynamo due to a turbulent flow, with the kinetic energy
spectrum $E\propto k^{-s}$ (where $k$ is the wave number), produce magnetic structures of length $l_0$ that
have a typical width of $l_0\Rm^{-1/2}$ and thickness $l_0\Rm^{-2/(1+s)}$ \citep{Wilkin2007}.
The magnetic correlation length scale $l_b$ discussed in Section~\ref{crpropg} ($l_b/l_0\simeq0.02$ in our simulations of the kinematic dynamo)
lies between the three characteristic scales. \citet{Wilkin2007} find that 
$l_b/l_0\simeq\Rm^{-0.4}$.
\citet{Sub99} suggests that the statistically steady (saturated) state of the dynamo corresponds to the
effective value of the magnetic Reynolds number $\Rm\simeq\Rmc$, where $\Rmc\simeq10^2$--$10^3$ is the 
critical magnetic Reynolds number for the dynamo action. Then $l_b/l_0\simeq\Rmc^{-0.4}\simeq
0.2\text{--}0.05$. Given that the magnetic field strength within such structures is close to
energy equipartition with the turbulent energy, the root-mean-square magnetic field strength (averaged
over a volume $l_0^3$, or larger) follows as 
$b_2 \simeq(l_b/l_0)^{3/2}\Beq\simeq \Rmc^{-0.6}\Beq\simeq(0.02\text{--}0.06)\Beq$
with $\Beq=(4\pi\rho u_0^2)^{1/2}\simeq5\mkG$. 
Numerical simulations suggest that the fluctuation dynamo produces a stronger rms magnetic field, 
$b_2\simeq0.3\Beq$ \citep{BS2005} in a saturated state of non-linear dynamo. This implies that random magnetic fields 
outside the filaments and ribbons contribute significantly to the magnetic energy density. 

Another contribution to non-Gaussian magnetic fields in the ISM is due to shocks. Primary and secondary shock fronts produced by supernova explosions and 
strong stellar winds can be described as pervasive shock-wave turbulence in the interstellar medium \citep{ByT87}. 
The typical separation between random shocks in the warm interstellar medium is $10^{16}\text{--}10^{17}\cm$ \citep{B88}. 
The magnetic field associated with shock-wave turbulence has a spectrum close to $k^{-2}$ and 
is intermittent at scales smaller than the separation of the shock fronts.
It is reasonable to expect that the energy density of these random magnetic fields is of the same order of 
magnitude as the kinetic energy of turbulence, $b_3\simeq\Beq$.

Overall, the small scale magnetic field in the ISM is a combination 
of both the non-Gaussian (due to fluctuation dynamo and shocks) and Gaussian (due to tangling of mean field) components of comparable 
energy density. Thus, we consider both the intermittent and randomized field for our study of cosmic ray propagation.

The magnetic field structure is likely to be different in different phases of the ISM. The warm, partially ionized
gas occupies a large fraction of the galactic disc volume and hosts both the large-scale and small-scale dynamos. The
gas has a scale height larger than the disk thickness and flows up to fill the hot galactic corona. Numerical simulations
of the multiphase ISM driven by supernovae suggest that the large-scale magnetic field is 
stronger in the warm phase whereas
random fields are present in all the phases of the ISM \citep{EGSFB17}. Also, scattering of cosmic rays due to self-generated
waves depends on the wave damping mechanisms, which differs in different phases of ISM \citep{CK81,FK2001}. 
The damping is strongest in the cold phase and is weakest in the hot phase (leading to efficient 
pitch angle scattering of cosmic rays).
So, the structure of the magnetic field and the self-generated waves would vary depending on the phase of the ISM. 

\section{Magnitude of magnetic fluctuations at the Larmor scale due to self-generated waves} \label{fluc}
When the  velocity of cosmic rays is greater than the local \Alfven speed, 
the streaming instability excites Alfv\'en waves of a wavelength
comparable to the particles' Larmor radius \citep{Wentzel1974,KP1969}.
Cosmic rays (originally propagating at speeds very close to the
speed of light) are slowed down by pitch angle scattering till they move around with the local \Alfven speed. We assume that the cosmic rays,
moving initially with the speed of light, transfer all of 
their momentum to \Alfven waves (no damping processes are assumed). 
Using conservation of momentum we can then estimate the maximum possible amplitude (i.e. $\delta B$) of these waves. 

The initial momentum of cosmic rays is
\begin{equation} \label{initialmom}
	p = n\cra mv \simeq n\cra mc \simeq \epsilon\cra/c\,,
\end{equation}
where $n\cra$ is the number density of cosmic ray particles, $m$ is the 
proton mass, $v$ is the particle speed (taken to be close to $c$, the speed of 
light) and $\epsilon\cra$ is the energy density of cosmic rays.
When all the momentum has been transferred to Alfv\'en waves, the
bulk speed of the ensemble of cosmic ray particles reduces to the Alfv\'en 
speed. Momentum conservation then implies that
\begin{equation} \label{finalmom}
	p = \frac{\delta B^2}{8\pi}\frac{1}{V\Al} + n\cra m V\Al\,,
\end{equation}
where $\delta B$ is the amplitude of the Alfv\'en waves generated by the streaming
instability, and $V\Al$ is the local Alfv\'en speed.
Combining Eqs.~\eqref{initialmom} and \eqref{finalmom}, we obtain, for $V\Al\ll c$,
\begin{equation} \label{equate}
 \epsilon\cra \simeq \frac{\delta B^2}{8\pi}\frac{c}{V\Al}\,.
\end{equation}
Assuming that the energy density of cosmic rays is comparable to the energy density 
of the magnetic field at larger scales, 
$\epsilon\cra \simeq B^2/(8\pi)$, it follows that
\begin{equation} \label{equate2}
 \frac{\delta B}{B} \simeq \left (\frac{V\Al}{c} \right)^{1/2}.
\end{equation}
The pitch angle scattering is most efficient in the hot phase of the ISM 
\citep{KP1969,CK81,Kulsrud2005}, where $B \simeq 5 \mu$G and 
$n \simeq 10^{-3}$cm$^{-3}$ for the thermal gas number density, 
the \Alfven speed is of the order of $10^{7}\cm\s^{-1}$.
Equation~\eqref{equate2} then yields $\delta B/B \simeq 10^{-2}$. 
Since the damping of Alfv\'en waves \citep{KP1969,Kulsrud2005} is neglected,
this is an upper estimate.

A similar estimate can be obtained without direct appeal to the energy equipartition
between the magnetic fields at a larger scale and cosmic rays. \cite{FG2004}, 
using properties of MHD turbulence, obtain $\delta B/B \simeq (\Lar/l_0)^{1/4}$,
where $\Lar$ is the Larmor radius of the particle and $l_0$ is the outer scale
of turbulence ($\sim 100 \pc$ for the ISM). 
For a $5 \GeV$ proton in a $5 \mkG$ magnetic field, the Larmor radius of the particle is $10^{12} \cm$.
Thus, at that scale, $\delta B/B \simeq (10^{12}/10^{18})^{1/4}\simeq 10^{-2}$, which agrees with our estimate
in the previous paragraph.
\begin{figure*} \centering
	\includegraphics[width=\columnwidth]{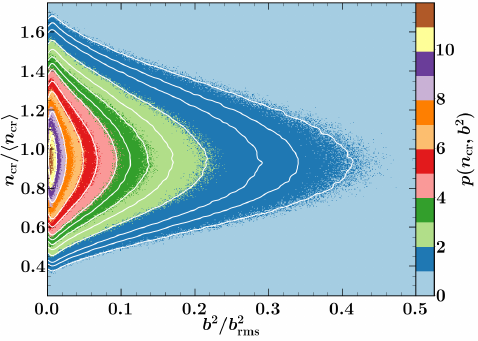}
	\includegraphics[width=\columnwidth]{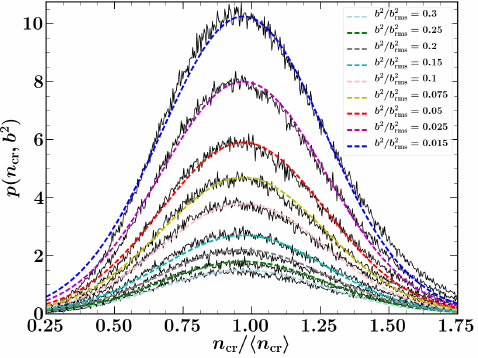}
	\includegraphics[width=\columnwidth]{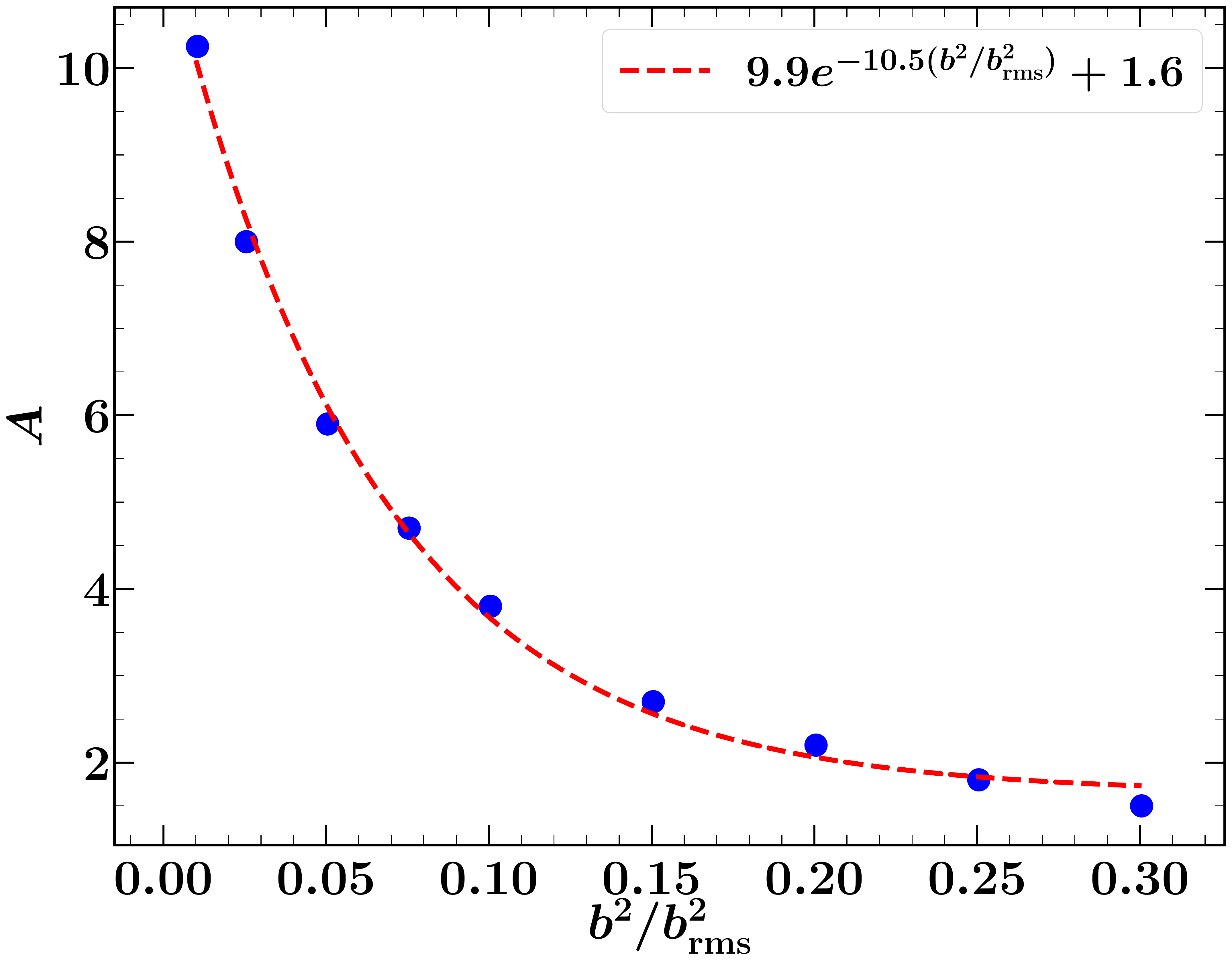}
	\includegraphics[width=\columnwidth]{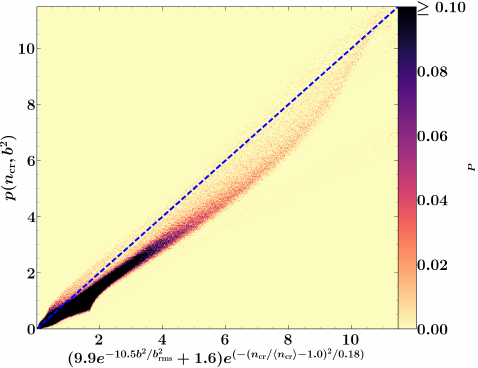}
	\caption{The joint probability density function (PDF) $p(\ncr,b^2)$
		of $b^2/\brms^2$ and $\ncr/\ncravg$  in the intermittent magnetic
		field for $0 \le b^2/\brms^2 \le 0.5$ and 
		$0.25 \le \ncr/\ncravg \le 1.75$ (upper left panel). The
		joint PDF $p(\ncr,b^2)$ as a function of $\ncr/\ncravg$
		alone for various fixed values of $b^2/\brms^2$ is
		shown in the upper right panel together with least-square
		fits of the form $p(\ncr,b^2)=A(b^2) \e^{-(\ncr-1.0)^2/0.18}$
		(smooth dashed curves).
		The lower left panel presents $A$ as a function of
		$b^2/\brms^2$ fitted with an exponential (red, dashed). 
		The lower right panel shows the $2d$ histogram of the
		computed and fitted values of $p(\ncr,b^2)$, with a
		linear fit (dashed, blue).
	}
	\label{probint}
\end{figure*}

\begin{figure*} \centering
	\includegraphics[width=\columnwidth]{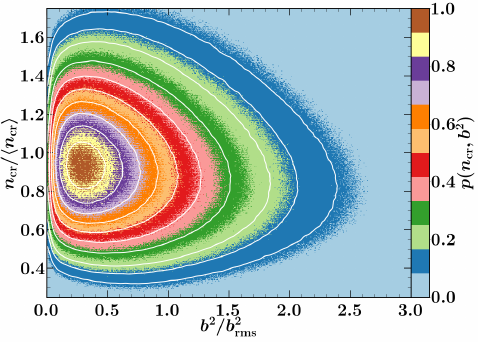}
	\includegraphics[width=\columnwidth]{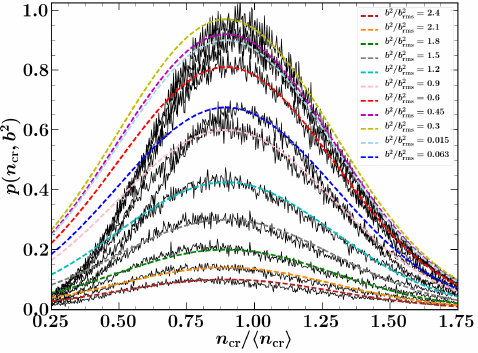}
	\includegraphics[width=\columnwidth]{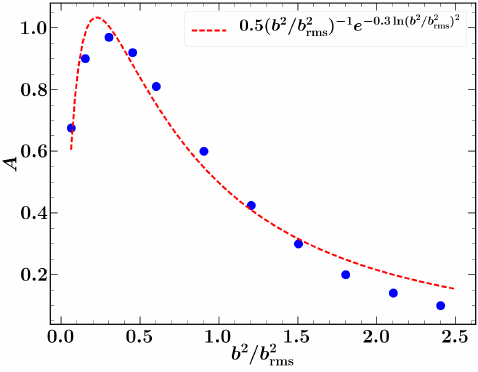}
	\includegraphics[width=\columnwidth]{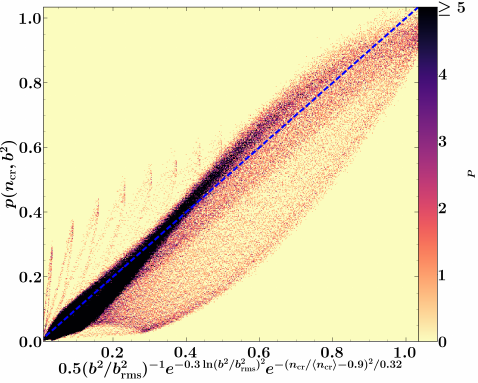}
	\caption{As in Fig.~\ref{probint} but for the randomized 
		(Gaussian) magnetic field for $0 \le b^2/\brms^2 \le 3.0$ and
		$0.25 \le \ncr/\ncravg \le 1.75$ and the lower right panel shows the $2d$ histogram with 
                blue-dashed line as $y=x$ (the bisector of the angle between the axes). 
	}        
	\label{probG}
\end{figure*}
\section{Statistical independence of cosmic ray and magnetic field distributions} \label{indep}

The distributions of cosmic rays and magnetic field energy density are uncorrelated (as discussed in Sec.~\ref{corr}) and we
check for their statistical independence, i.e., whether $p(\ncr, b^2)$, the joint
probability distribution function of two components is equal to $p(\ncr)p(b^2)$, the product
of the probability distribution function of each component. 
This is tested especially for the core region where $\ncr$ is close to its mean value. 
The upper left panels of Figures~\ref{probint} and \ref{probG} show the joint probability 
densities of magnetic field strength and cosmic ray number densities
in the intermittent and Gaussian (randomized) magnetic fields,
respectively. We use the Kolmogorov--Smirnov (KS) test to check
whether the two distributions are independent. We perform KS 
test on marginal probability distributions, $p(b|\ncr)$
and $p(\ncr|b)$ to check whether each of them are drawn
from the same distribution. This is measured by the $D$ statistics of
the test, which is the absolute maximum distance between the cumulative
distribution function of the two samples.
A value of $D$ smaller than 0.087 represents $95\%$ confidence
that the two samples are drawn from the same distribution.
For this we consider two samples from the marginal probability distribution   
$p(b|\ncr)$ with consecutive $\ncr$ (cuts along consecutive values of 
$y$ axis of upper left panel of \Fig{probint}). We performed the KS test for
all such pairs and found the mean value of $D$ to be $\simeq 0.062$.
A similar calculation for $p(\ncr|b)$ gives $\simeq0.072$. 
The mean $D$ values for Gaussian (randomized) magnetic fields are
$\simeq0.054$ for $p(\ncr|b)$ and $\simeq0.114$ for $p(b|\ncr)$ respectively.
The small $D$
values suggest that the cosmic ray number
density and magnetic field distributions are independent of each other
and the joint probability $p(\ncr, b^2)$ can be factorized into 
$p(\ncr)$ and $p(b^2)$. We find the functional dependence of 
$p(\ncr)$ on $\ncr$ and $p(b^2)$ on $b^2$ as follows.
Cuts through the joint PDF in the intermittent magnetic 
field, shown in the upper right panel of Fig.~\ref{probint}
are well fitted with a Gaussian which, remarkably, has neither the mean
value nor the standard deviation dependent on $b$. The dependence
of the maximum value of the conditional PDF $p_n(\ncr|b^2)$ on $b^2$
is shown in the lower left panel of Fig.~\ref{probint} together with its
fit with an exponential function.  Thus, the joint PDF of $\ncr$
and $b$ has the form
\begin{equation}\label{fit_pi}
	p(\ncr,b)\approx(1.6+9.9
		\e^{-10.5 b^2/\brms^2}) \e^{-(\ncr/\langle\ncr\rangle-1)^2/0.18}
\end{equation}
for $0 \le b^2/\brms^2 \le 0.3$ and $0.25 \le \ncr/\ncravg \le 1.75$.
The relative accuracy of the fitted parameters is better than 
$5$ per cent.
This is confirmed by the $2d$ histogram of the measured and fitted
values of $p(\ncr,b)$ shown in the lower left panel of 
Fig.~\ref{probint}: for a perfect fit, the points would be all on
the bisector of the quadrant angle. The scatter about that 
line, shown dashed (blue) provides a measure of the accuracy of the fit.
A very small number of points systematically deviate from the main dependence,
yet lying on a straight line as well. The joint PDF is thus factorizable
emphasizing that both the distribution are independent. 
 
Similar analysis for $p(\ncr,b)$ in the randomized (Gaussian)
magnetic field is illustrated in \Fig{probG}. As in the intermittent
magnetic field, the joint PDF is roughly factorizable and thus $\ncr$ and $b$ are
statistically independent. The form of the conditional PDF of
magnetic field strength is a lognormal distribution,
\begin{equation}\label{fit_pbG}
	p(b|\ncr)\approx 0.5 (b^2/\brms^2)^{-1}\e^{-0.3 \ln(b^2/\brms^2)^2}
\end{equation}
for $0\leq b^2/\brms^2\leq3$. We find that 
\begin{equation}\label{fit_piG}
	p(\ncr,b)\approx 0.5 (b^2/\brms^2)^{-1}\e^{-0.3 \ln(b^2/\brms^2)^2} \e^{-(\ncr/\langle\ncr\rangle-0.9)^2/0.32}\,.
\end{equation}
The $2d$ histogram of the measured and fitted values is shown (with the dashed (blue)
line for the perfect match) in the lower right panel of \Fig{probG}.

To summarize, the distributions of cosmic ray number density and
magnetic field strength are statistically independent in the diffusive 
regime of the cosmic rays. In the intermittent magnetic field, the joint 
PDF of $\ncr$ and $b$ is well approximated by a Gaussian in $\ncr$
and a modified Gaussian in $b$. In randomized (Gaussian) field, the joint
PDF is approximated by a Gaussian in $\ncr$ and a lognormal distribution in $b^2$.
The two variables remain statistically independent for cosmic rays of other energies too. 
The exact parameters of the joint PDF are likely to depend on details of the magnetic field structure
and the energy of the particle. 

\label{lastpage}
\end{document}